\documentclass[a4paper,11pt]{article}
\pdfoutput=1
\usepackage{jheppub}
\usepackage{amsmath}
\usepackage{amsthm}
\usepackage[normalem]{ulem}
\usepackage{amsfonts}
\usepackage{amssymb}
\usepackage{textcomp}
\usepackage{bm}
\usepackage{color}
\usepackage{verbatim}
\usepackage{subcaption}
\usepackage{graphicx,color}
\usepackage{epsfig}

\definecolor{orange}{rgb}{1,0.5,0}
\definecolor{col1}{RGB}{153, 52, 121}
\definecolor{dgreen}{rgb}{0,0.55,0}
\definecolor{pink}{rgb}{1,0.08,0.58}

\usepackage{amsfonts,amsmath,amssymb}

\newcommand{\cool}{\ensuremath{%
  \mathchoice{\includegraphics[height=2ex]{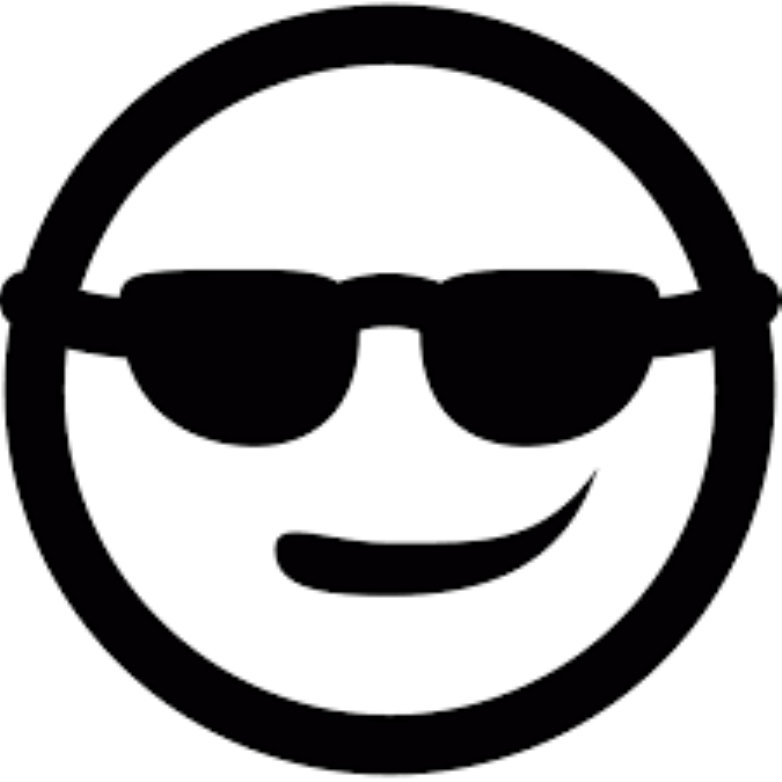}}
    {\includegraphics[height=2ex]{cool.pdf}}
    {\includegraphics[height=1.5ex]{cool.pdf}}
    {\includegraphics[height=1ex]{cool.pdf}}
}}

\newcommand{\la}{\langle}
\newcommand{\ra}{\rangle}

\newcommand{\rar}{\rightarrow}

\theoremstyle{definition}

\usepackage[normalem]{ulem}

\begin{document}

\subheader{CCTP-2017-5\\ ITCP-IPP 2017/16}

 \title{
 \Huge Pinning of longitudinal phonons\\ in holographic spontaneous helices \color{black}}
 
 \author[a]{Tomas Andrade}
 
 \author[\cool]{Matteo Baggioli} 
  
 \author[b,c]{Alexander Krikun\footnote{https://orcid.org/0000-0001-8789-8703}}
 
  \author[b]{Napat Poovuttikul}

 \affiliation[a]{Rudolf Peierls Centre for Theoretical Physics \\ University of Oxford, 1 Keble Road, Oxford OX1 3NP, UK} 

 \affiliation[\cool]{Crete Center for Theoretical Physics, Institute for Theoretical and Computational Physics, Department of Physics, University of Crete, 71003
Heraklion, Greece} 

 \affiliation[b]{Instituut-Lorentz, Universiteit Leiden \\ P.O. Box 9506, 2300 RA Leiden, The Netherlands} 
 \affiliation[c]{Institute for Theoretical and Experimental Physics (ITEP)\footnote{On leave from}, \\ B. 
 Cheryomushkinskaya 25, 117218 Moscow, Russia }

\emailAdd{tomas.andrade@physics.ox.ac.uk}
\emailAdd{mbaggioli@physics.uoc.gr}
\emailAdd{krikun@lorentz.leidenuniv.nl}
\emailAdd{poovuttikul@lorentz.leidenuniv.nl}

\abstract{We consider the spontaneous breaking of translational symmetry and identify the associated Goldstone mode -- a longitudinal phonon -- in a holographic model with Bianchi VII helical symmetry. 
For the first time in holography, we observe the pinning of this mode after introducing 
a source for explicit breaking compatible with the helical symmetry of our setup. 

We study the dispersion relation of the resulting pseudo-Goldstone mode, uncovering how its speed 
and mass gap depend on the amplitude of the source and temperature. 
In addition, we extract the optical conductivity as a function of frequency, which reveals a metal-insulator transition as a consequence of the pinning. 
}

\maketitle

\section{Introduction}
 
Translation symmetry breaking is ubiquitous in any condensed matter system, the ionic lattice playing a fundamental role in almost 
every model attempting to describe realistic matter. 
From the point of view of the electronic subsystem, the lattice is usually considered as an external, explicit source of translation symmetry breaking. 
On the other hand, many strongly coupled electronic systems tend to break translations spontaneously, forming superstructures unrelated to the underlying ionic lattice. These include charge density waves, spin density waves, various intertwined orders. 

The spontaneous symmetry breaking always comes together with the appearance of a gapless mode in the spectrum, the Goldstone boson, associated to the generator of the broken symmetry. 
In the case of the spontaneously formed atomic crystal, these are known as acoustic phonons, while in the context of the charge density wave the Goldstone is usually referred to as a gapless sliding mode.
If translations are not broken explicitly, momentum is still conserved and the gapless mode mediates transport, thus ensuring 
that the currents overlapping with momentum do not decay so that the corresponding DC conductivity is infinite\footnote{An elegant way of obtaining this result is given by the memory matrix formalism \cite{Hartnoll:2012rj,Lucas:2015vna}. See also \cite{Nitta:2017mgk} for a recent
field theoretic discussion of relativistic Goldstone modes associated to the breaking of translations.}. 
The situation changes when the breaking of translations is explicit, due to an atomic lattice or quenched disorder being taken into account. In this case, generically, the Goldstone boson acquires a mass proportional to the strength of the explicit symmetry breaking parameter and turns into a pseudo-Goldstone (the $\pi$-meson, Goldstone of the chiral symmetry breaking, being a familiar example of this effect). 
More concretely, when the explicit breaking scale is small in comparison to the spontaneous symmetry breaking scale, one can write down the leading terms in a gradient expansion of the dispersion relation of the would-be-Goldstone boson as
\begin{equation}\label{pseudo}
\omega^2\,= - i \omega \Gamma + \omega_0^2 + c_s^2 q^2+...
\end{equation}
where $\omega_0$ is the \textit{pinning frequency}, or mass gap, $\Gamma=\tau_{rel}^{-1}$ is the damping parameter present because of momentum relaxation, $q$ is the momentum of the phonon and $c_s$ its speed. Note that the gradient expansion is valid when 
$\{\omega_0, \Gamma, \omega,  q \} \ll T$. The ellipsis denotes higher order dissipative terms due to various viscosities. Moreover, 
the phonon behaves as a long-lived excitation when $\omega_0 \gg \Gamma$. It is in this regime when we can treat it in the hydrodynamic
approximation in a well-defined way.
In absence of spatially dependent sources, both the mass gap and damping are zero and one recovers the usual massless phonon with linear dispersion relation. 
The pinning frequency is therefore proportional to the scale of explicit symmetry breaking. Given that the existence of the Goldstone itself relies on the spontaneous symmetry breaking, it should also be expected that the pseudo-Goldstone mass is proportional to the order parameter of spontaneous symmetry breaking as well.\footnote{In case of the $\pi$-meson this is taken into account by the Gell-Mann--Oaks--Renner relation: $M_\pi^2 \sim m_q \sigma$, where $\sigma$ -- the chiral condensate -- is the order parameter of the spontaneous chiral symmetry breaking, while $m_q$ -- the quark mass -- breaks chiral symmetry explicitly.}

The dynamics of the pseudo-Goldstone can be further quantified in the framework of a gradient expansion. Before going further, let us summarize the results obtained in models where translations are broken in one direction only, which will help us to analyze the numerical results which 
we will obtain later on. More specifically, we will regard the formulae coming from the 1D models as phenomenological, e.g. we are not assuming that the various parameters that enter in these formulae have a precise mathematical definition in our theory. Instead, we will take them 
as a qualitative guidance to interpret our numerical results. As we shall see, all the qualitative expectations that follow from the 1D models 
are in good agreement with our results.

The nature of the gapped and damped phonon is directly reflected in the low frequency structure of the optical conductivity.  Our focus is the longitudinal optical conductivity where the electric field is applied in the direction parallel to the direction of broken translational symmetry. This conductivity, in the presence of the charge density $\rho$ takes the following form  \cite{Delacretaz:2016ivq,Delacretaz:2017zxd}\footnote{In the case of Galilean invariance, $\sigma_0=0$, the electric conductivity $\sigma$ and related expressions have been already derived in
\cite{RevModPhys.60.1129}. }
\begin{equation}\label{cond}
\sigma(\omega) = \sigma_0 + \frac{\rho^2}{\chi_{\pi\pi}}\, \left(\frac{- i \omega}{- i \omega (\Gamma - i \omega) + \omega_0^2}\right),
\end{equation}
and can be extracted in the framework of gradient expansion which we briefly discuss in appendix \ref{app-hydro}. The first term $\sigma_0$ is an independent first order transport coefficient which appears in the constitutive relation for the current $J^\mu$\
More information on the transport of the system can be obtained by turning on a perturbation with finite wave vector $q^i$ and study the 
momentum resolved  current-current correlator.
In this work, we consider the case where the 
propagation direction is parallel to the single direction in which translational symmetry is broken.
It turns out that the pole of such two-point function is governed by the mode with the dispersion relation \eqref{pseudo} and, in absence of explicit symmetry breaking, the propagation velocity of this longitudinal mode encodes the information regarding the elastic modulus in the following schematic form
\begin{equation}\label{speeds}
c_s^2 = c_0^2 + \frac{\kappa}{\chi_{\pi\pi}},
\end{equation}
where $c_0$ is a function of thermodynamic quantities $\chi_{\pi\pi}$ is the momentum susceptibility\footnote{In the framework of \cite{Delacretaz:2017zxd}, the momentum susceptibility is defined as $T^{ti} = \chi_{\pi\pi} v^i$, where $T^{\mu\nu}$ is the stress-energy tensor and $v^i$ is the fluid velocity. In the relativistic fluid $\chi_{\pi\pi}=\varepsilon+p$,  where $\varepsilon$ and $p$ are the energy density and pressure of the system, respectively.}
and $\kappa$ is the elastic modulus. Close to the phase transition $\kappa$ vanishes and $c_0$ is the speed of sound in the translational invariant medium.

This expression conveys the fact that once translations are spontaneously broken, the speed of sound acquires additional contributions which encode the elastic effects of a given material. Note, however, that the formula \eqref{speeds} is derived for a system with broken translational symmetry in one direction as discussed in e.g. \cite{Delacretaz:2017zxd} which does not take into account the helical structure in the example we considered\footnote{It would be desirable to incorporate the symmetry of the helix in the constitutive relation and derive e.g. the speed of sound and the Kubo formulae for the elastic moduli. We plan to revisit this issue in the near future.}.
Nevertheless, \eqref{speeds} gives a good qualitative understanding of the physics around the phase transition and we will use it as 
a practical tool to compare with our numerical results.
In particular, we will see that the speed of sound in our model has a kink below $T_c$, in good qualitative agreement with 
the physics of a second order phase transition.

In case of the charge density wave with explicit symmetry breaking, the appearance of 
a finite gap in the spectrum of the sliding mode means that it is ``pinned'' and cannot freely propagate any more. 
This pinning introduces a gap in momentum transport and the system undergoes a metal-insulator phase transition. 

In this paper we study the pinning of Goldstone modes in a strongly coupled system which has a holographic dual, taking advantage 
of the fact that generic properties of strongly coupled phases in which translational symmetry is broken can be conveniently accessed 
in this class of models \cite{Zaanen:2015oix}. 
This endeavour amounts to the construction of black holes which lack translational symmetry along the boundary 
directions, either spontaneously by condensation, explicitly by the choice of the boundary conditions, or in a mixture of both mechanisms.
As we shall describe in detail below, we are able to reproduce the above expectations regarding the nature of the pseudo-Goldstone bosons
summarized in \eqref{pseudo}
-\eqref{speeds}.

Among the studies of purely spontaneous structures in holography we can mention the formation of helices \cite{Nakamura:2009tf, Ooguri:2010kt, Donos:2012wi}, charge density waves \cite{Donos:2013gda}, helical superconductors \cite{Donos:2011ff} and various types of striped orders \cite{Donos:2011bh, Donos:2011pn, Donos:2011qt, Rozali:2012es, Donos:2013woa, Withers:2013kva, Withers:2013loa, Jokela:2014dba, Withers:2014sja, Krikun:2015tga, Erdmenger:2013zaa, Cremonini:2016rbd, Ling:2014saa, Cai:2017qdz}. The generic feature is that, as first noted in the pioneering work of \cite{Nakamura:2009tf}, below a certain critical temperature the Reissner-Nordstr\"om (RN) solution exhibits near horizon instabilities with finite momentum which indicate the emergence of a new branch of spatially modulated ground states. 

Holographic models of explicit symmetry breaking have also been widely considered. These include the study of metallic, insulating and superconducting phases alongside with transitions amongst them 
\cite{Donos:2012js, Donos:2013eha, Andrade:2013gsa, Donos:2014oha, Donos:2014uba, Gouteraux:2014hca, Taylor:2014tka, Horowitz:2012ky, Horowitz:2013jaa, Donos:2014yya, Rangamani:2015hka, Erdmenger:2015qqa, Andrade:2014xca, Kim:2015dna, Ling:2014laa, Ling:2014saa, Baggioli:2015zoa, Baggioli:2015dwa}. As a result of the explicit breaking of translations, the DC conductivity is finite in all these models.

More recently, the interplay of the explicit and spontaneous breaking has been investigated in the holographic context \cite{Andrade:2015iyf, Andrade:2017leb, Jokela:2016xuy, Cremonini:2017usb}. The key aspect of these references is that there are distinct mechanisms (and in particular, bulk fields) which trigger the spontaneous breaking and allow for the introduction of explicit breaking, so, we can separately dial off the explicit breaking while retaining the physics of the spontaneous structure. 

All of the above models come in two broad classes, which can be identified depending on whether the breaking of translations occurs in a generic way or along some particular compensating symmetry direction. The former configurations are usually termed as ``inhomogeneous", and at the technical level they are characterized by the fact that the relevant equations of motion are PDEs. 
%
On the other hand, the latter are known as ``homogeneous", and the resulting equations are ODEs\footnote{At the operational 
level, these share many similarities with the Massive Gravity model introduced in \cite{Vegh:2013sk}, as explained in \cite{Blake:2013owa}.}. 
%
%
Remarkably, the simplified homogeneous models capture most of the relevant physics, e.g. 
finite DC conductivities in the presence of explicit symmetry breaking, although the appearance of commensurability effects requires the system to be inhomogeneous \cite{Andrade:2015iyf, Andrade:2017leb}.

Here we will work with a model which belongs to the homogeneous class -- the Bianchi VII helix model \cite{Ooguri:2010kt,Nakamura:2009tf,Donos:2012js,Donos:2012wi,Donos:2014oha,Ammon:2016szz}.
The distinctive feature of this setting is a particular ``helical'' pattern of the symmetry breaking\footnote{It would be interesting to analyse this system from the coset construction point of view similar to \cite{Nicolis:2013lma} to have a better understanding of the symmetry breaking pattern and the corresponding gapless excitations.}. Bianchy VII helix is realized in the model with 3+1 dimensional boundary theory. In case of 3 spacial dimensions ($x$,$y$ and $z$) one can separate two \textit{commuting} symmetry transformations: translation along the $x$-axis (with generator $P_x$) and rotation in the transverse $(y,z)$-plane (with generator $J_{yz}$). When the helix (with pitch $k$) is formed, the direct product of these two symmetries is broken down to the diagonal subgroup. While the normal state is invariant under the action of either $P_x$ or $J_{yz}$, in the broken state only the combination $H_+^k = P_x + k J_{yz}$ remains a symmetry. In case of the spontaneous symmetry breaking, one expects one Goldstone mode to arise due to the broken orthogonal generator $H_-^k = P_x - k J_{yz}$. This Goldstone mode has nonzero projection on $P_x$, therefore we expect to find its manifestation in the spectrum of conductivity, which overlaps with momentum.

The substantial technical simplifications, which we enjoy in the helical model are due to the remaining diagonal symmetry subgroup, generated by $H_+^k$. It allows us to rewrite the equations of motion as ODEs, making the analysis much easier. 
The price we pay is the fact that translational symmetry breaking in this model is in practice unidirectional along $P_x$. Moreover, the technical advantages of the model would be immediately lost if one 
considered propagation across the helix director, which would break the helical symmetry generated by $H_+^k$. Hence, we will restrict our considerations to the longitudinal phonons propagating along the helix:  a direction in which the translational symmetry is broken.

We will construct numerically the fully back-reacted black holes
which break translations spontaneously and obtain their spectra of quasi-normal 
modes (QNM), which via the standard holographic dictionary \cite{Son:2002sd} maps to the 
set of poles of the retarded correlators of the dual field theory. When the symmetry breaking is
purely spontaneous, we will find a Goldstone mode which we identify with the phonon. 
Upon the introduction of a small explicit breaking, we will confirm 
that the mode becomes pinned, i.e. its dispersion relation being modified according to \eqref{pseudo}.\footnote{Also in the holographic context, some authors have considered solutions that explicitly break translational invariance and interpreted the fluctuations of the corresponding structure as pseudo-Goldstone modes \cite{Baggioli:2014roa,Baggioli:2015gsa, Alberte:2015isw, 
Argurio:2016xih, Amoretti:2016bxs}.} 
In addition, we will consider the optical conductivity of the corresponding black holes, and verify that at low frequencies
it can be well approximated by \eqref{cond}. Moreover, we will observe that the pinning 
induces a metal-insulator transition due to the shift of the Drude peak to higher frequencies.\footnote{It should be emphasized that the state arising due to this transition cannot strictly speaking be classified as an insulator since it lacks 
an insulating gap\cite{Grozdanov:2015qia}. Instead, it belongs to the generic holographic class of ``algebraic insulators'' where the conductivity vanishes at zero temperature as a power law, without exponential suppression due to the finite gap. This particular feature will not play a 
significant role in the present study.}
Furthermore, as we lower the temperature below the phase transition, we observe that the speed of sound $c_s$ increases continuously from its translational invariant value, revealing a non-zero elasticity modulus\footnote{Previous discussions about the elastic properties in holography can be found in \cite{Alberte:2015isw,Alberte:2016xja} in the context of massive gravity theories.} as can be seen from \eqref{speeds}.

This paper is organized as follows: In Section \ref{sec:the_model} we introduce our holographic model, and study the background solution and the set of relevant perturbations. In Section \ref{sec:Goldstone} we analyze in detail the spectrum of QNMs and confirm the theoretical expectations within our concrete model. Finally, in Section \ref{sec:discussion} we conclude discussing our results and presenting some possible future directions. In appendix \ref{app-hydro}, we summarise the results from a hydrodynamic-like effective theory and further elaborate on how they can be applied to our holographic setup. 
In appendix \ref{app-linear}, we provide the technical details involving how to construct the linearised perturbation equations. In \ref{app-num}, we provide more explanations about the numerical techniques involved in the computations.

\section*{\sc Note added}
While this work was being finished, Refs.\cite{Jokela:2017ltu,Alberte:2017cch}, which studied interplay between explicit and spontaneous symmetry breaking in different holographic models,  were also being completed and appeared on the arXiv on the same day as our article. 

\section{The holographic model}
\label{sec:the_model}

We consider the 5D setup of \cite{Nakamura:2009tf}, which consists of an Einstein-Maxwell-Chern-Simons model. 
As shown in \cite{Donos:2012wi}, the system admits fully non-linear solutions which break translations purely spontaneously 
respecting Bianchi VII helical symmetry of given momentum~$k$. 
We break translations explicitly as in \cite{Donos:2012js} via the introduction of a helical source of amplitude $\lambda$ for an additional vector field. 
For simplicity, we take the momentum of the explicit helix to be equal to the spontaneous momentum $k$, 
since this allows us to reduce all equations of motion to ODEs.  
It should be noted that the resulting configurations will 
generically not be preferred in the grand canonical ensemble. To see this, we recall from the results of \cite{Andrade:2015iyf} that helical lattices do not display commensurability effects in the structure of the instabilities towards the formation of spontaneous symmetry breaking, 
so that it is unlikely that the pitches of the explicit and spontaneous structures will lock at the non-linear level. However, we expect these effects to 
be negligible for small enough lattice amplitude, so our present analysis should indeed be adequate to shed light on the mechanism of pinning of the Goldstone modes associated to translational invariance.

Denoting the gauge field by $A$ and the additional vector field by $B$, with the strengths $F= dA$ and $W = dB$, we write our action as
\begin{equation}
 	S = \int d^5 x \sqrt{- g} \left( R  - 2 \Lambda - \frac{1}{4} F^2 - \frac{1}{4} W^2  \right) - 
 	 \frac{\gamma}{6} \int d^5 x A \wedge F \wedge F
\end{equation} 
We fix the AdS radius to unity be setting $\Lambda = - 6$. In the following, we shall consider $\gamma = 1.7$
to ease comparison with previous results in the literature.
A similar model was considered in \cite{Erdmenger:2015qqa}, which focused on the condensation of a charged scalar instead 
of the spontaneous breaking of translations. The UV asymptotics and relevant holographic renormalization 
results can be read off from \cite{Erdmenger:2015qqa} and the earlier reference \cite{Donos:2012wi}.

%

\subsection{Background solutions}
\label{sec:background}

The configurations we study are of the form
\begin{align}
\label{anstaz ds2}
ds^2 &= \frac{1}{u^2} \left[ - Q_{tt} f dt^2 + \frac{Q_{uu} du^2}{f} + Q_{11} (\omega^{(k)}_1)^2 + Q_{22} ( \omega^{(k)}_2 +  Q_{t2} dt) ^2 
	+ Q_{33} (\omega^{(k)}_3)^2 \right] \\
\label{ansatz A B}
	A &=  A_t dt +  A_2 \omega^{(k)}_2 , \qquad B =  B_t dt + B_2 \omega^{(k)}_2
\end{align}
All the unknowns $Q_{tt}$, $Q_{uu}$, $Q_{11}$, $Q_{22}$, $Q_{33}$, $Q_{t2}$, $A_t$, $A_2$, $B_t$, $B_2$ 
are functions of the single variable $u$, which corresponds to the holographic coordinate. 
We take the basis of the helical structure to be
\begin{align}
\omega^{(k)}_1 & = dx \\
\omega^{(k)}_2 & = \cos (k x) dy - \sin(k x) dz \\
\omega^{(k)}_3 & = \sin (k x) dy + \cos(k x) dz 
\end{align}

It is worth mentioning that the gauge is not fully fixed in our ansatz, which can be seen by noting that the equations of motion 
do not determine all the unknown functions. We will fix this redundancy shortly.

We can write the RN solutions as  $Q_{tt} = Q_{uu} = Q_{11} = Q_{22} = Q_{33} = 1$, $Q_{t2} = A_2 = B_t = B_2 = 0$ and 
\begin{equation}
	A_t = \mu(1 - u^2), \qquad f = (1- u^2)\left( 1 + u^2 - \frac{1}{3} \mu^2 u^4 \right)
\end{equation}
We have chosen coordinates such that the conformal boundary and horizon are located at $u=0$ and $u=1$, respectively.
We will construct more general finite temperature solutions by imposing regularity of all the unknown functions at the horizon.
In the UV, we demand the metric to be AdS in addition to the following conditions for the gauge fields
\begin{equation}
	A_t = \mu + O(u^2) , \qquad A_2 = O(u^2), \qquad B_t = O(u^2), \qquad B_2 = \lambda + O(u^2)
\end{equation}
As usual, we interpret $\mu$ as the chemical potential.
In the case $B= 0$, our solutions will reduce to the non-linear spontaneous helices of \cite{Donos:2012wi}, in which the spontaneous current
can be read off from the subleading term in $A_2$.
In order to break translational invariance explicitly without directly sourcing the current sitting in $A_2$, we will turn 
on the source $\lambda$ in the gauge field $B$. When doing so, consistency of the equations of motion requires a non-vanishing $B_t$ component 
when $Q_{t2} \neq 0$. We set the leading term in $B_t$ to zero to avoid introducing extra sources. We emphasize that the 
gauge field $B$ is only a technical device to introduce explicit breaking of translations. The conserved current associated to the chemical 
potential, and the one we shall be concerned with when computing transport properties, is the one sitting in $A$\footnote{In 
principle one could add a mass $m_B^2$ for the vector field $B$, which removes the extra conserved current without changing 
the qualitative picture in the metric-$U(1)$ sector. We set $m_B^2 = 0$ only to simplify the UV asymptotics. A similar strategy 
was adopted in \cite{Erdmenger:2015qqa}.}.

We will construct our solutions by starting with a spontaneous helical structure of momentum $k$, and place it on an explicit lattice
{\it of the same momentum} by turning on $\lambda$. As mentioned in the Introduction, this keeps the equations of motion as ODEs 
thus largely simplifying our calculations. We solve the non-linear equations of motion by using the DeTurk trick 
\cite{Headrick:2009pv,Adam:2011dn,Wiseman:2011by}, in combination with a Newton-Raphson relaxation 
method with pseudospectral collocation. The DeTurk method then fixes the aforementioned gauge symmetry of the metric in a 
dynamical way\footnote{In principle, the system of ODEs one obtains from Einstein and Maxwell's equations can be solved using shooting 
(after fixing the gauge properly, setting e.g. $Q_{33} =1$), but in order to achieve the desired accuracy to study the fluctuations, 
we find it technically convenient to employ a relaxation method since the output of the method includes the values of the 
unknowns at the ends of the computational domain and no regulators are introduced on the boundaries, as it is necessary 
for the shooting method to work \cite{Hartnoll:2009sz}. Relaxation has previously been used to construct similar 
helical backgrounds in \cite{Andrade:2015iyf, Bagrov:2016cnr}.
}.
In this gauge, the equations of motion imply that $Q_{tt}= Q_{uu}$ at the horizon $u=1$, which in turn fixes the ratio 
\begin{equation}
	\frac{T}{\mu} = \frac{2}{3} \frac{6 - \mu^2}{4 \pi \mu}
\end{equation}
Our backgrounds are characterized by four parameters, which we take to be the temperature $T$, the momentum $k$, the 
lattice amplitude $\lambda$ and the chemical potential $\mu$. Due to the underlying conformal symmetry, we can 
scale away one of these parameters. We choose instead to express all physical quantities in units of the chemical potential. 

\begin{figure}[ht]
\centering
\includegraphics[width=0.6 \linewidth]{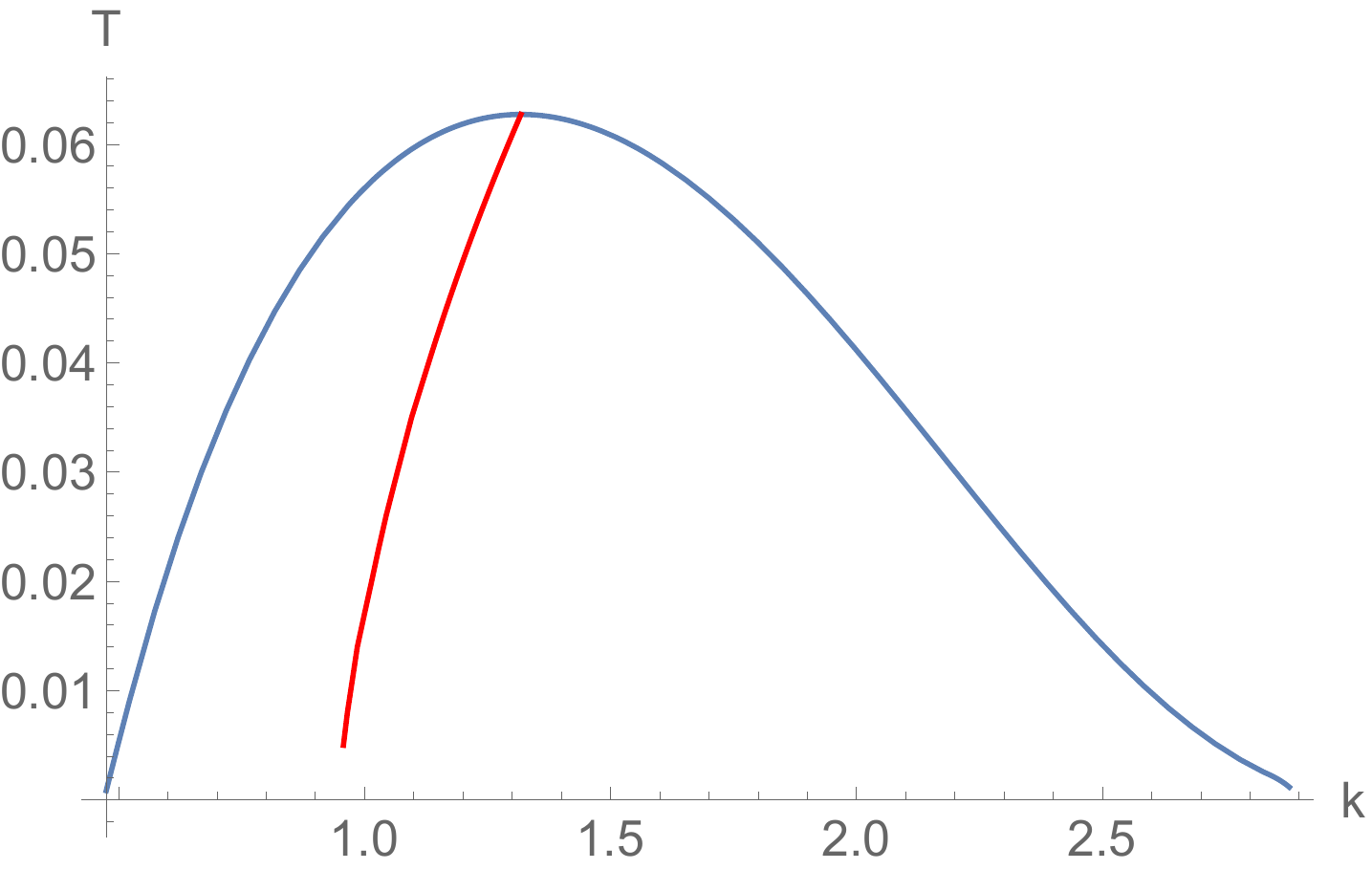}
\caption{\label{fig:bell} Marginal modes (blue) and thermodynamically preferred solutions for 
$\gamma = 1.7$. Our results are in agreement with \cite{Donos:2012wi}}
\end{figure}

In the absence of sources, our solutions reduce to the ones constructed in \cite{Donos:2012wi}. These are characterized by 
the temperature and momentum of the spontaneous helix, and populate the interior of the ``bell-curve" determined by the marginal 
modes. For a given temperature, there is one solution inside the bell-curve which minimizes the free energy, which is given 
by \cite{Donos:2012wi}
\begin{equation}
	w = \frac{1}{2}( T_{tt} - T_{xx} ), 
\end{equation}
\noindent where $T_{tt}$ and $T_{xx}$ are components of the (renormalized) boundary stress tensor.
The set of these solutions forms a curve in the plane $(k,T)$ which characterizes the thermodynamically preferred configurations
in the grand canonical ensemble, see Fig.\,\ref{fig:bell}. In the following we shall restrict ourselves to these solutions 
prior to the introduction of explicit symmetry breaking.

We are interested in the behaviour of these solutions as we turn on the spatially dependent source parametrized by $\lambda$. 
Above the critical temperature, the resulting configurations are the ones found in \cite{Donos:2012js}. These correspond
to solutions which break translations explicitly and, as such, posses finite DC conductivity. Their metal/insulator phase 
diagram was considered in \cite{Donos:2012js} and later expanded in \cite{Andrade:2015iyf}. In the present study we will 
consider the regime in $(\lambda,k)$ plane in which the high temperature solutions are metallic. Below the critical temperature, 
the configurations of interest correspond to the spontaneous structure placed on an external lattice of the same momentum. 
Note that the critical temperature at which the formation of the spontaneous structure occurs depends on $\lambda$ \cite{Andrade:2015iyf}.
It is worth mentioning that after turning on the spatially dependent source, some logarithmic terms appear in the 
boundary expansion \cite{Donos:2012js, Donos:2012wi, Erdmenger:2015qqa}, with the associated 
conformal anomaly \cite{Henningson:1998gx, deHaro:2000vlm}. For our purposes, this presents only a mild technical issue and does not influence our main results. 
Moreover, we have checked that our numerical results are not very sensitive to these log terms, 
in that it suffices to use pseudospectral method to reach adequate convergence. 

We also note that the solutions which we are considering, are fine tuned in a sense that the periodicity of the spantaneous and explicit structures coincide. Moreover, in \cite{Andrade:2015iyf} it has been already shown that, due to the lack of commensurability effects in this type of models, our solutions are not thermodynamically preferred for sizable $\lambda$. This feature may lead to the appearence of the unstable mode in the spectrum, corresponding to the change of the period of the spontaneous helix. Nonetheless, it will not substantially affect the features of the Goldstone mode, which is in the focus of the present study.

\subsection{Linearized Perturbations and Goldstone modes}
\label{sec:perturbations}

Our main goal is to study the physical properties of the pseudo-Goldstone mode associated to 
the translational symmetry along the $x$ direction. At zero lattice amplitude $\lambda = 0$, 
this mode becomes a true Goldstone and can be identified by acting with a translation on the background
configuration\footnote{The existence of this mode follows simply by diffeomorphism invariance 
and the translational invariance of the background configuration.}. By taking the Lie derivative along 
$\partial_x$ of  the ansatz \eqref{anstaz ds2}, \eqref{ansatz A B} with $B= 0$, we obtain the perturbation
\begin{align}\label{GM}
\delta (ds^2) &= - \frac{2 k}{u^2} \omega^{(k)}_3 \bigg[ dt Q_{t2} Q_{22} +  \omega^{(k)}_2 (Q_{22} - Q_{33})  \bigg] \\
\delta A &= k A_2 \omega^{(k)}_3
\end{align}
We can readily check that this mode is a solution of the linearized equations of motion, and gives the profile for the 
Goldstone mode at zero frequency $\omega$ and perturbation momentum~$q$\footnote{Strictly speaking, one 
should regard $q$ as a Bloch momentum, due to the breaking (either spontaneous or explicit) of translational invariance.
However, because of the ODE nature of the equations of motion, there is no Brilloiun zone and this distinction does not play
any significant role.}. In order to obtain the dispersion relation for this mode, however, 
we need to include the dependence $\sim e^{- i \omega t + i q x}$, which couples it to {\it all} the remaining fluctuations
of the metric $\delta g_{\mu \nu}$ and gauge fields $\delta A_\mu$ and $\delta B_\mu$. 
We describe our gauge fixing procedure and other details of our linear mode calculation in appendix \ref{app:linear_eqs}.

We compute the conductivity by 
turning on the source for the linearized electric field, while keeping all the other sources to zero. This can be expressed in 
terms of the UV asymptotics of the perturbation 
$\delta A_x$, 
\begin{equation}
  	\delta A_x = e^{- i \omega t + i q x} 	( \delta A_x^{(0)} + u^2 \delta A_x^{(2)} + u^2 \log u  \delta \tilde A_x^{(2)}  + \ldots ) 
\end{equation} 
\noindent where $\delta A_x^{(0)}$, $\delta A_x^{(2)}$, $\delta \tilde A_x^{(2)}$ are constants. The electric field in momentum space is then given by  
$E_x  = i \omega \delta A_x^{(0)}$. By choosing the appropriate renormalization scheme, we can define the boundary current to be simply 
proportional to $\delta A_x^{(2)}$, with no extra contributions from the log term $\tilde \delta A_x^{(2)}$, see \cite{Erdmenger:2015qqa}, 
so the optical conductivity can be expressed as
\begin{equation}
\label{sigma_definition}
	\sigma(\omega) = \frac{ \delta A_x^{(2)} }{i \omega \delta A_x^{(0)}}
\end{equation}
We note that the ambiguity in the renormalization scheme does not affect the IR behaviour of this observable, since different
schemes are related to each other by local terms which are of order $O(\omega^2)$. 

The QNMs which give the poles of the corresponding two point functions are obtained by setting all the sources for the linearized perturbations to zero and solving the Sturm-Liuville problem. We will present our numerical results for the optical conductivity and QNMs in the next section.

\section{Dispersion relation of the (pseudo-)Goldstone and AC conductivity}
\label{sec:Goldstone} 

\begin{figure}[ht]
\center
\begin{minipage}{0.58 \linewidth}
\includegraphics[width=1.\linewidth]{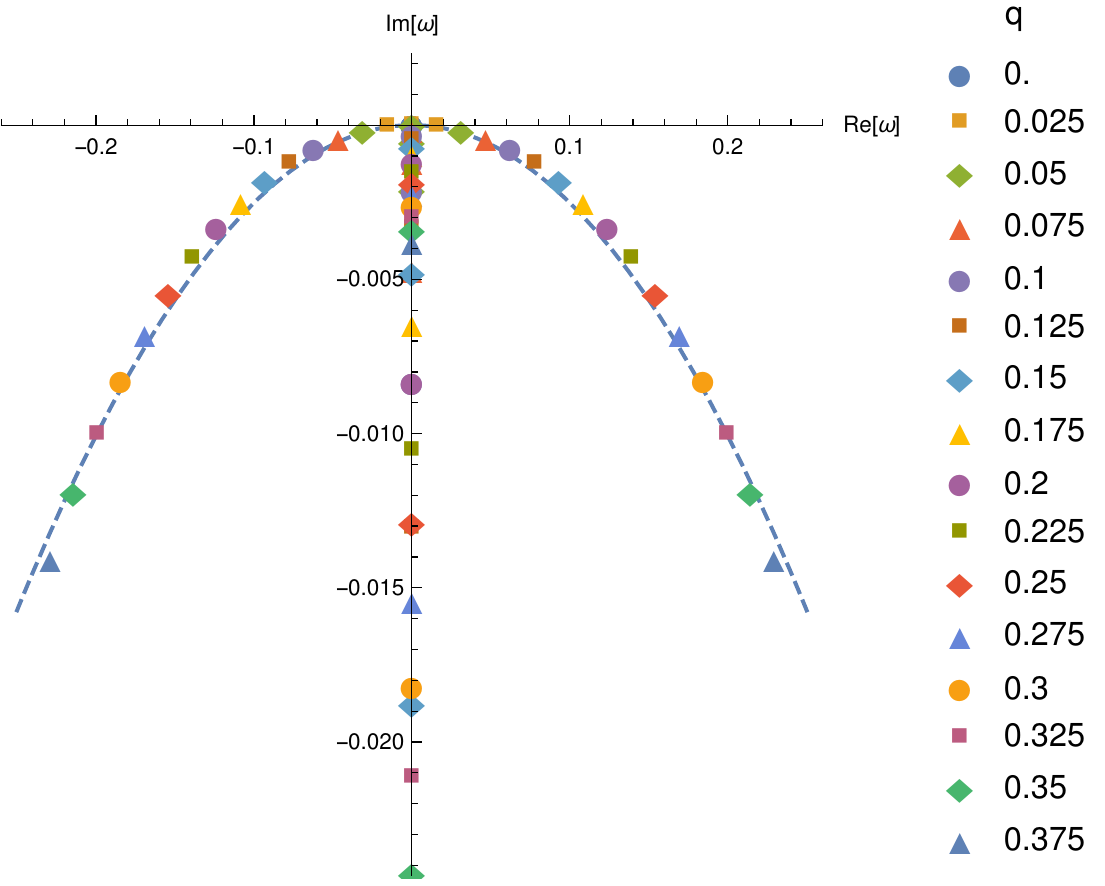}
\end{minipage}
\begin{minipage}{0.4 \linewidth}
\includegraphics[width=1.\linewidth]{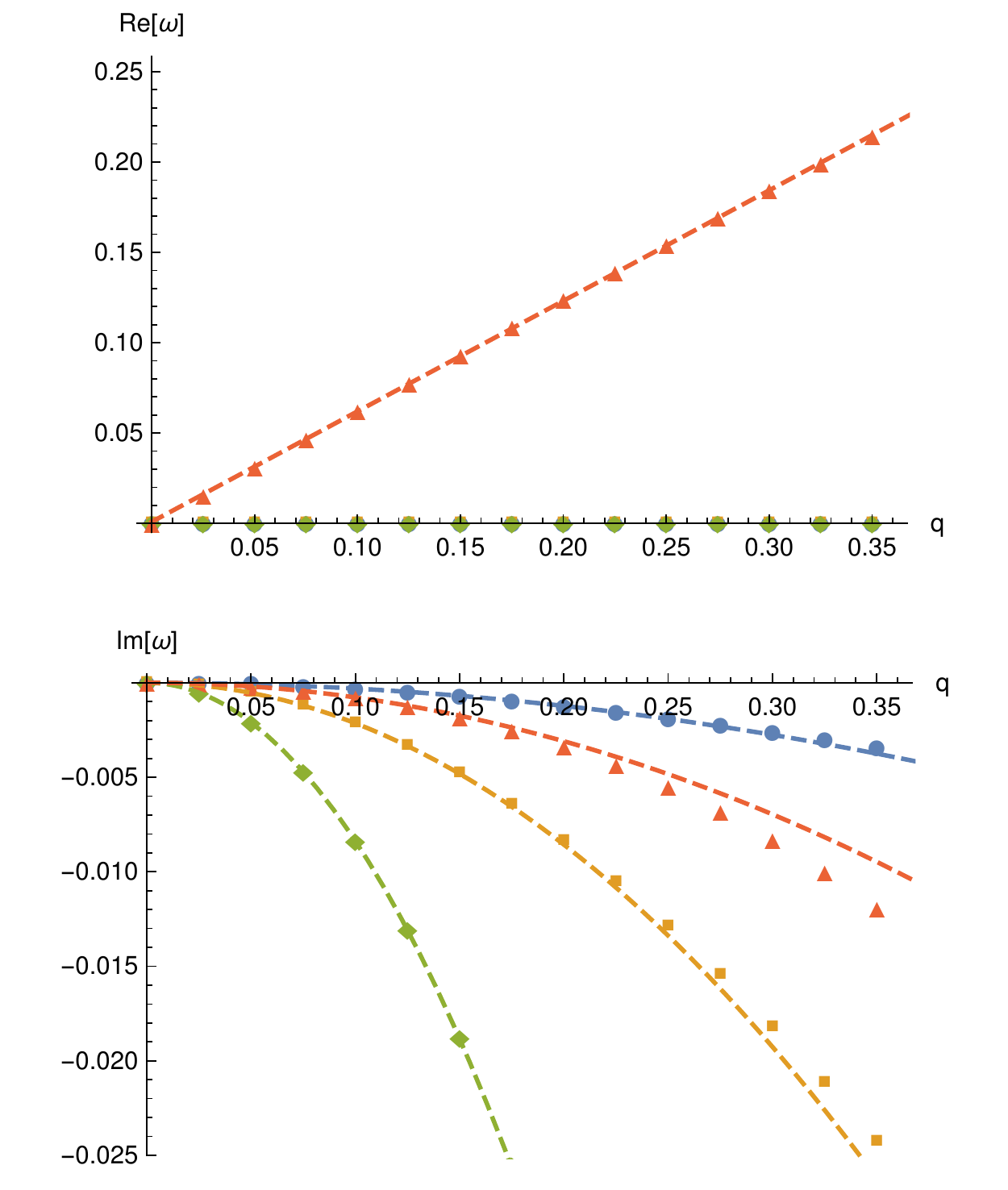}
\end{minipage}
\caption{\label{Fig:VaryQ_QNMs} Quasinormal modes in the purely spontaneously broken phase ($T=0.52 T_c$), showing among others the dispersion relation of the Goldstone mode (red dots on the right panel) as a function of momentum $q$. Left panel: trajectories of QNMs in the complex plane, Right panel: dependence of the real and imaginary parts on momentum.}
\end{figure}

We start our analysis by addressing the purely spontaneous symmetry breaking case $\lambda=0$. At finite temperature $T=0.0325$ well below the phase transition ($T_c = 0.0625$) we study the momentum dependent QNMs in the spectrum of the nonlinear spontaneous ground state (see Appendices \ref{app-linear}, \ref{app-QNM} for details of our numerical procedure.) The results are shown in Fig.\,\ref{Fig:VaryQ_QNMs}. 
One can readily discern the propagating gapless mode with linear dispersion relation -- the longitudinal phonon. The imaginary part of the QNM, 
related to the attenuation rate of the phonon, rises as $q^2$. Apart from the propagating mode we see 3 diffusive modes: the shear diffusive mode, the thermoelectric diffusive mode and the crystal diffusive mode. The existence of these low-energy excitation is in perfect agreement with the prediction from effective hydrodynamic-like theory, as we discuss in appendix \ref{app-hydro}.

\begin{figure}[ht]
\center
\includegraphics[width=0.5\linewidth]{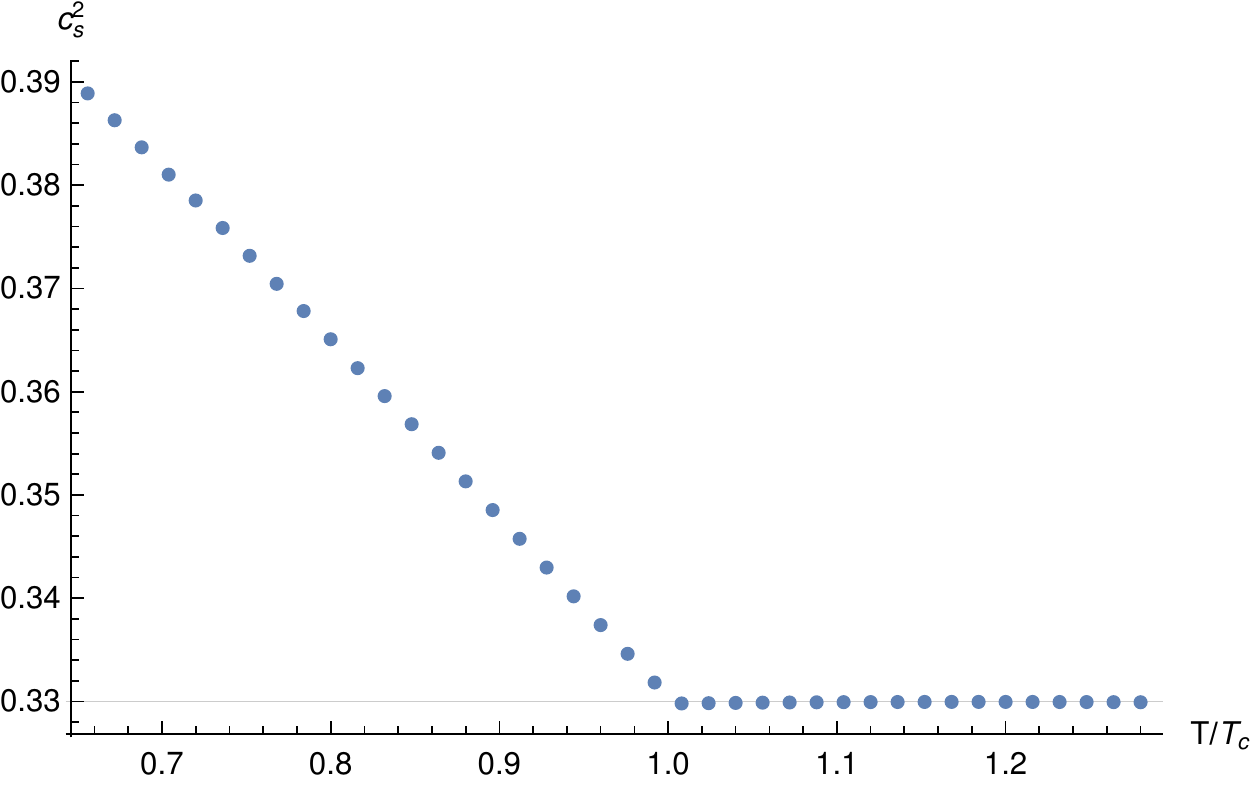}
\caption{\label{Fig:VsPlot} Dependence of the sound velocity squared on temperature.}
\end{figure}

It also is interesting to study the dependence of the speed of sound with temperature. For this purpose we evaluate the dispersion of phonon for a set of backgrounds with different temperatures and extract its slope near $q=0$, Fig.\,\ref{Fig:VsPlot}. Clearly, the square of the sound velocity rapidly rises below the critical temperature, as the bulk elastic modulus of the spontaneous structure increases, in perfect qualitative agreement with \eqref{speeds}.

\begin{figure}[t]
\center
\begin{minipage}{0.58 \linewidth}
\includegraphics[width=1.\linewidth]{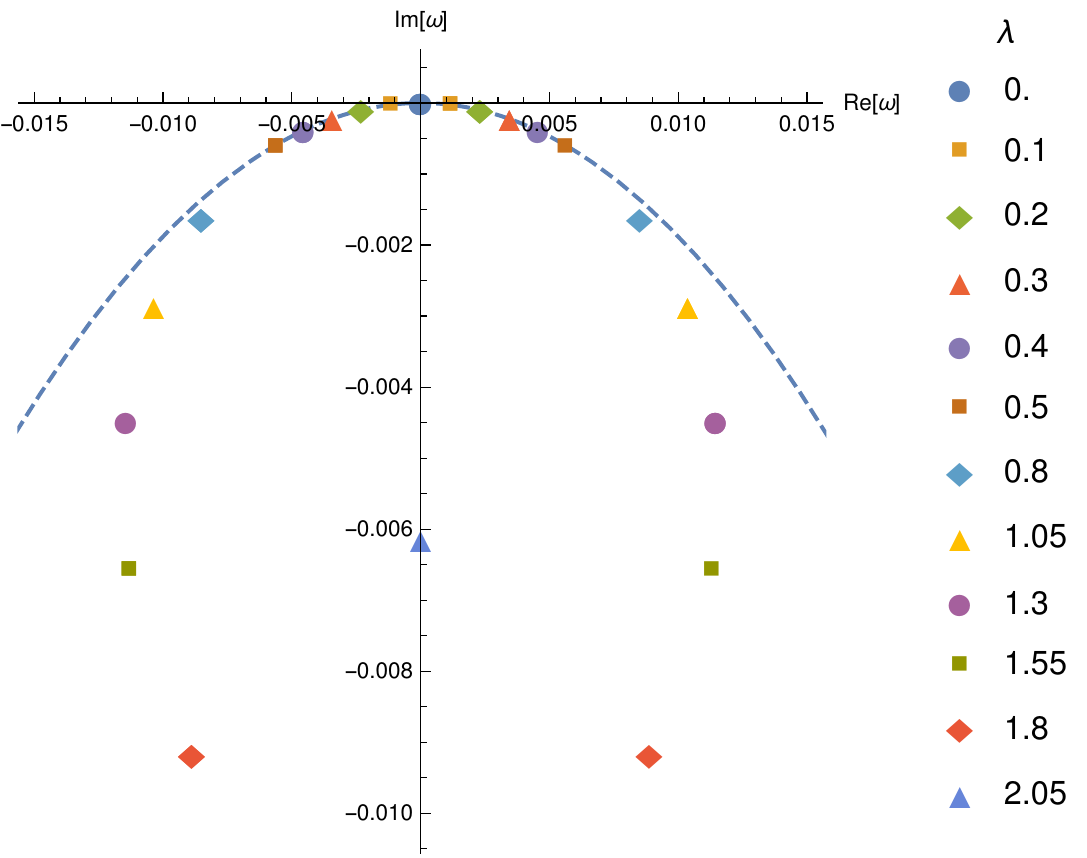}
\end{minipage}{}
\begin{minipage}{0.4 \linewidth}
\includegraphics[width=1.\linewidth]{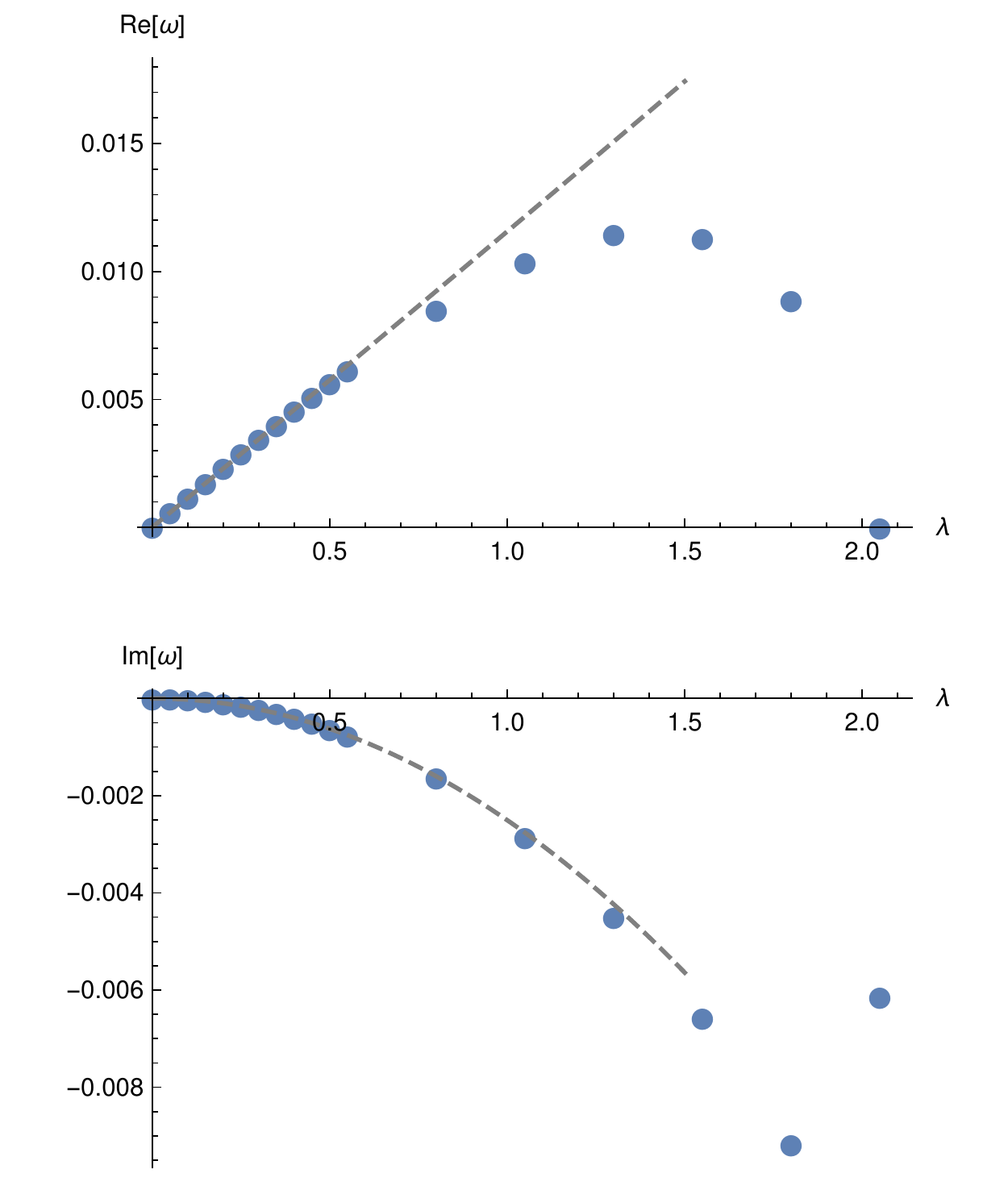}
\end{minipage}
\caption{\label{Fig:VaryL_QNMs} QNMs at zero momentum in the spontaneously broken phase ($T=0.52 T_c$)with increasing amplitude of explicit symmetry breaking 
$\lambda$. The rising pinning frequency of the phonon is seen (right panel) as a function of $\lambda$. Left panel: trajectories of QNMs in the complex plane, Right panel: dependence of the real and imaginary parts on $\lambda$.}
\end{figure}

If we now introduce explicit symmetry breaking, by slowly rising the source for the explicit helix $\lambda$, we will observe the phonon to acquire a finite pinning frequency, see Fig.\,\ref{Fig:VaryL_QNMs}. We observe that for small amplitude of the explicit helix $\lambda$, the pinning frequency $\omega_0$ is proportional to $\lambda$, while the width $\Gamma$ rises quadratically,
\begin{equation}
\label{equ:omega_lambda_scaling}
\omega_0 \sim \lambda, \qquad \Gamma \sim \lambda^2.
\end{equation}
This is in~accordance with the general expectations for a pseudo-Goldstone, outlined in the Introduction. The specific powers 
of $\lambda$ which appear may seem surprising, however, since the pinning frequency is proportional to the mass of the pseudo-Goldstone. In the simplest scalar field model, the mass enters the equation of motion squared and therefore one would expect $m^2$ to be proportional to the explicit symmetry breaking scale $\lambda$. But in the case under consideration, the field $B$, responsible for the explicit symmetry breaking, always enters the equations of motion quadratically. Therefore the effective scale of the symmetry breaking is $\lambda^2$ and \eqref{equ:omega_lambda_scaling} reflects this fact.

\begin{figure}[ht]
\center
\includegraphics[width=0.6\linewidth]{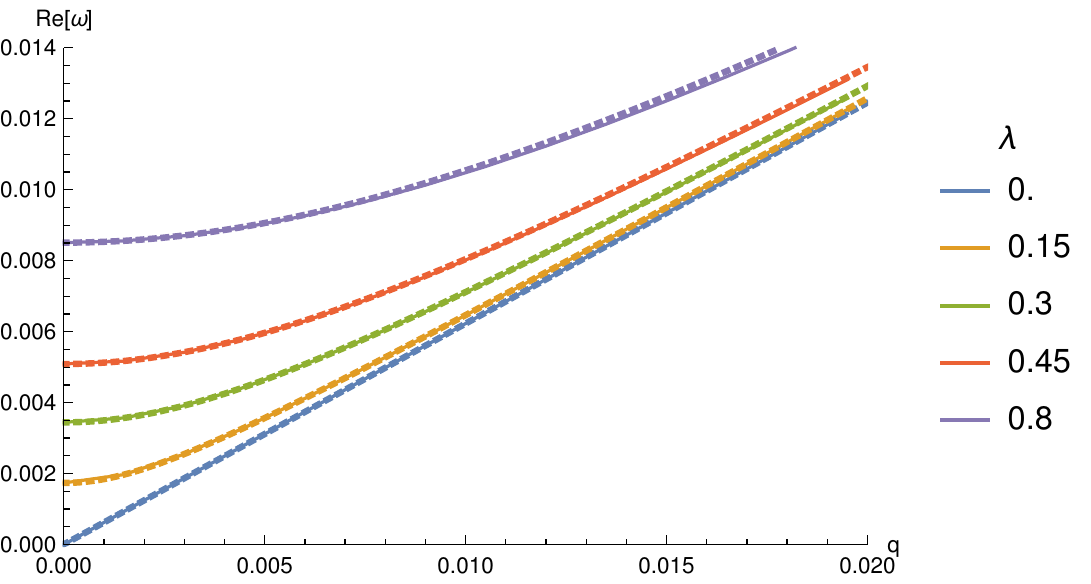}
\caption{\label{Fig:GapDispersion} Dispersion relation of the pinned phonon at different explicit lattice amplitudes. Solid lines show the numerical data for momentum dependent QNMs, dashed lines show the dispersion \eqref{pseudo} with parameters obtained from QNMs at zero momentum, Fig.\,\ref{Fig:VaryL_QNMs} and $c_0$ extracted from $\lambda=0$ dispersion, Fig.\,\ref{Fig:VaryQ_QNMs}.}
\end{figure}

Using the obtained values of $\omega_0$ and $\Gamma$, and the sound velocity, evaluated for purely spontaneous case, we can directly check the validity of the expression \eqref{pseudo} for the dispersion relation. On Fig.\,\ref{Fig:GapDispersion} we plot the real part of the momentum dependent QNM of the pseudo-Goldstone mode overlaid with the prediction of \eqref{pseudo}. Excellent agreement is apparent. 

Interestingly, the behaviour of pseudo-Goldstone QNMs at large $\lambda$ is qualitatively different. Starting from $\lambda \approx 1$ the gap 
no longer increases and the QNMs approach the imaginary axis. In this regime, the scale of explicit symmetry breaking becomes comparable with the scale of spontaneous symmetry breaking and hence the treatment of pseudo-Goldstone mode as a slight modification of the pure Goldstone 
does not apply. 

\begin{figure}[ht]
\center
\begin{minipage}{0.5 \linewidth}
\includegraphics[width=1.\linewidth]{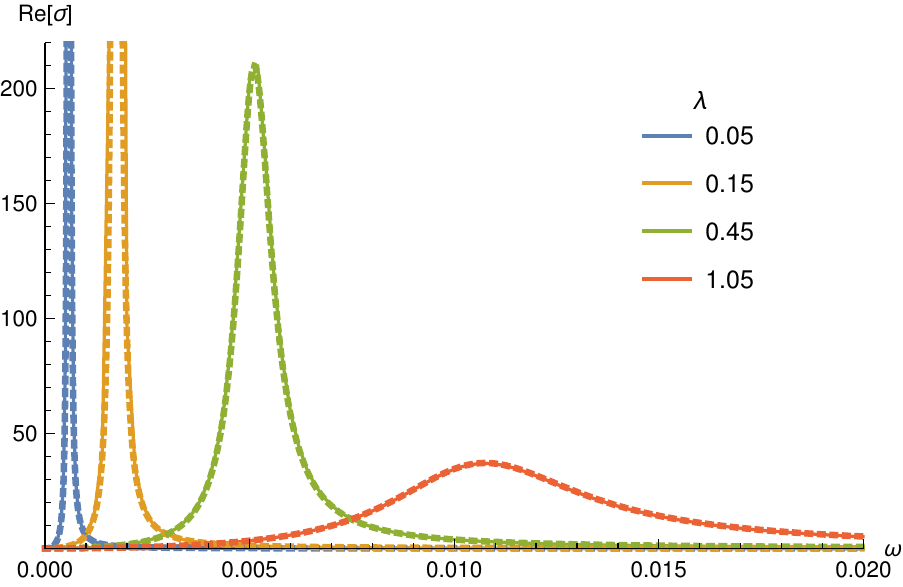}
\end{minipage}{}
\begin{minipage}{0.49 \linewidth}
\includegraphics[width=1.\linewidth]{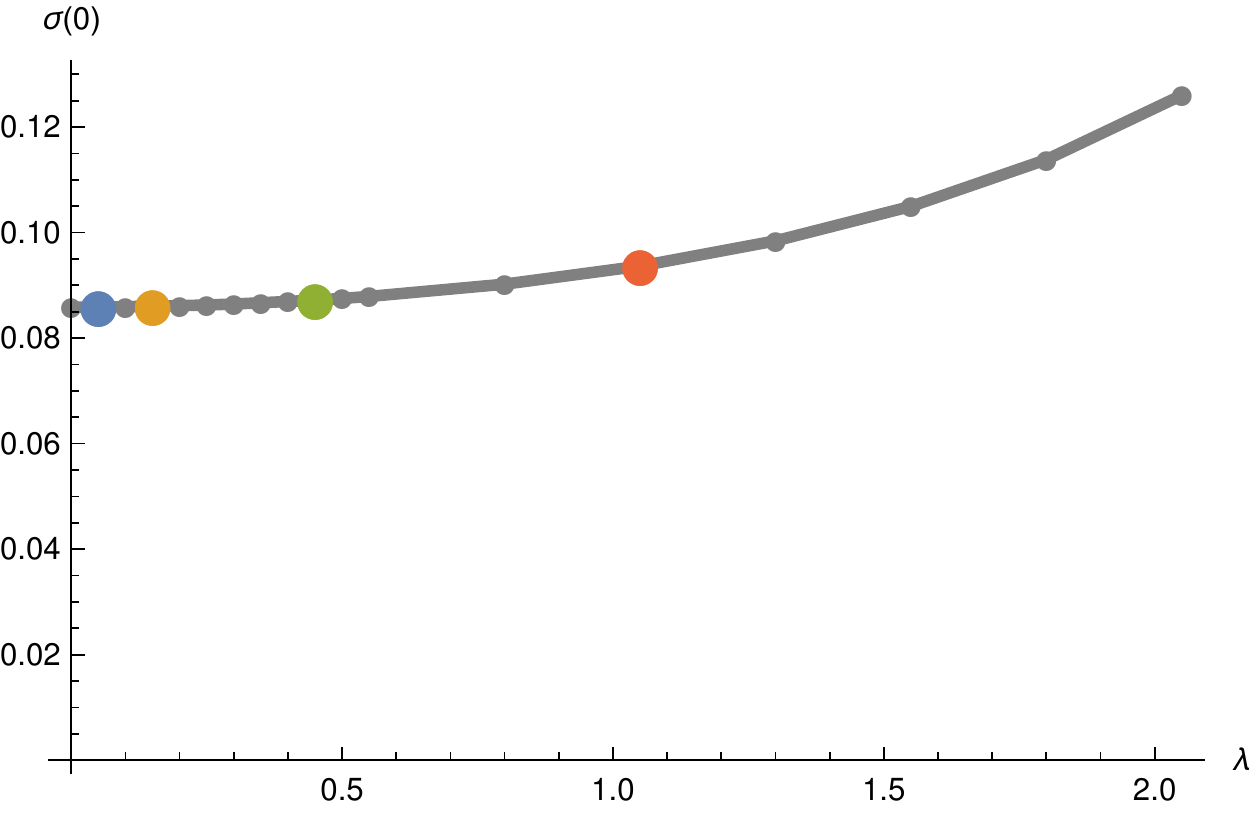}
\end{minipage}
\caption{\label{Fig:LamAC} AC conductivity (Left panel) and its limit at zero frequency (Right panel) in the state with fully developed spontaneous structure ($T=0.52 T_c$) as at different amplitudes of the explicit helix $\lambda$. The dashed lines show the prediction of \eqref{cond} with parameters extracted from QNM spectrum. The solid lines are the numerical results. Color dots on the right panel mark the parameters for which the AC curves has been produced. }
\end{figure}

We can further investigate the dependence of the AC and DC conductivity on $\lambda$ in the regime well below the critical temperature, where the spontaneous structure is fully developed. On the left panel of Fig.\,\ref{Fig:LamAC} we plot AC conductivity profiles for several values of $\lambda$ together with the prediction which follows form the gradient expansion formula \eqref{cond} with parameters obtained from the 
QNMs study. The only value which is fitted is the overall factor in front of the pole term, and this turns out to be constant for all $\lambda$ in the scope. We observe excellent agreement even in the extreme regime were explicit symmetry breaking scale is very small. For any finite value of $\lambda$, the Drude peak is pinned and moves to higher frequency, leaving zero contribution to the DC conductivity at $\omega=0$. This leads to a spectacular effect shown on the right panel of Fig.\,\ref{Fig:LamAC}, where we plot the limiting value (see Appendix \ref{app-QNM})
\begin{equation}
\sigma(0) \equiv \lim_{\omega \rar 0} \sigma(\omega).
\end{equation}
Even though the conductivity at $\lambda=0$ must be infinite due to the translation symmetry, the pinning of phonon with arbitrary small strength drives it down to $\sigma_0$ from \eqref{cond}, gapping away all the spectral weight contained in the Drude peak. Since all the available weight has been abruptly gapped, at $\lambda \lesssim 1$ the DC conductivity remains constant with $\lambda$. 
We do observe that this dependence grows slightly in the region of larger $\lambda$, where as we discussed above, 
the pseudo-Goldstone logic fails and the validity of \eqref{cond} is lost.

\begin{figure}[t]
\center
\begin{minipage}{0.64 \linewidth}
\includegraphics[width=1.\linewidth]{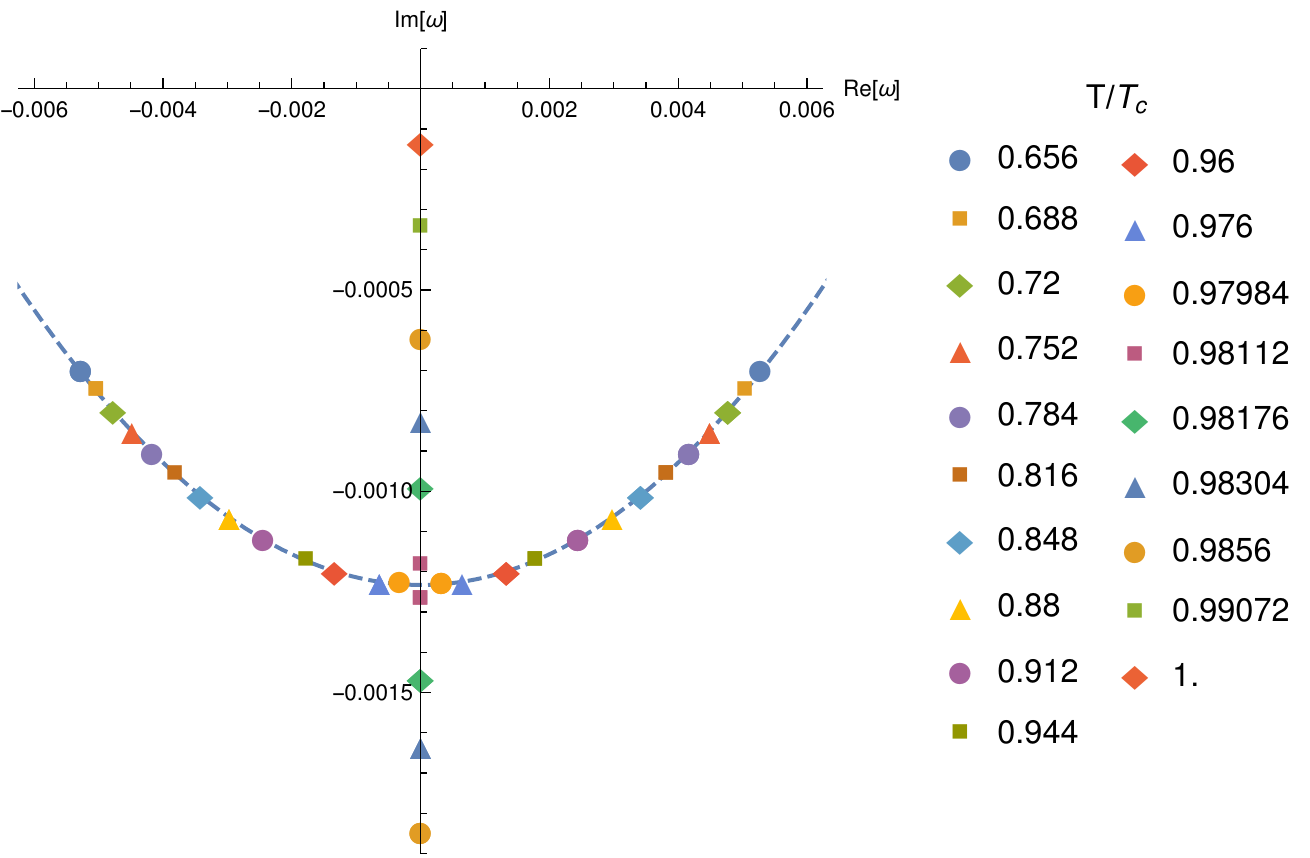}
\end{minipage}{}
\begin{minipage}{0.35 \linewidth}
\includegraphics[width=1.\linewidth]{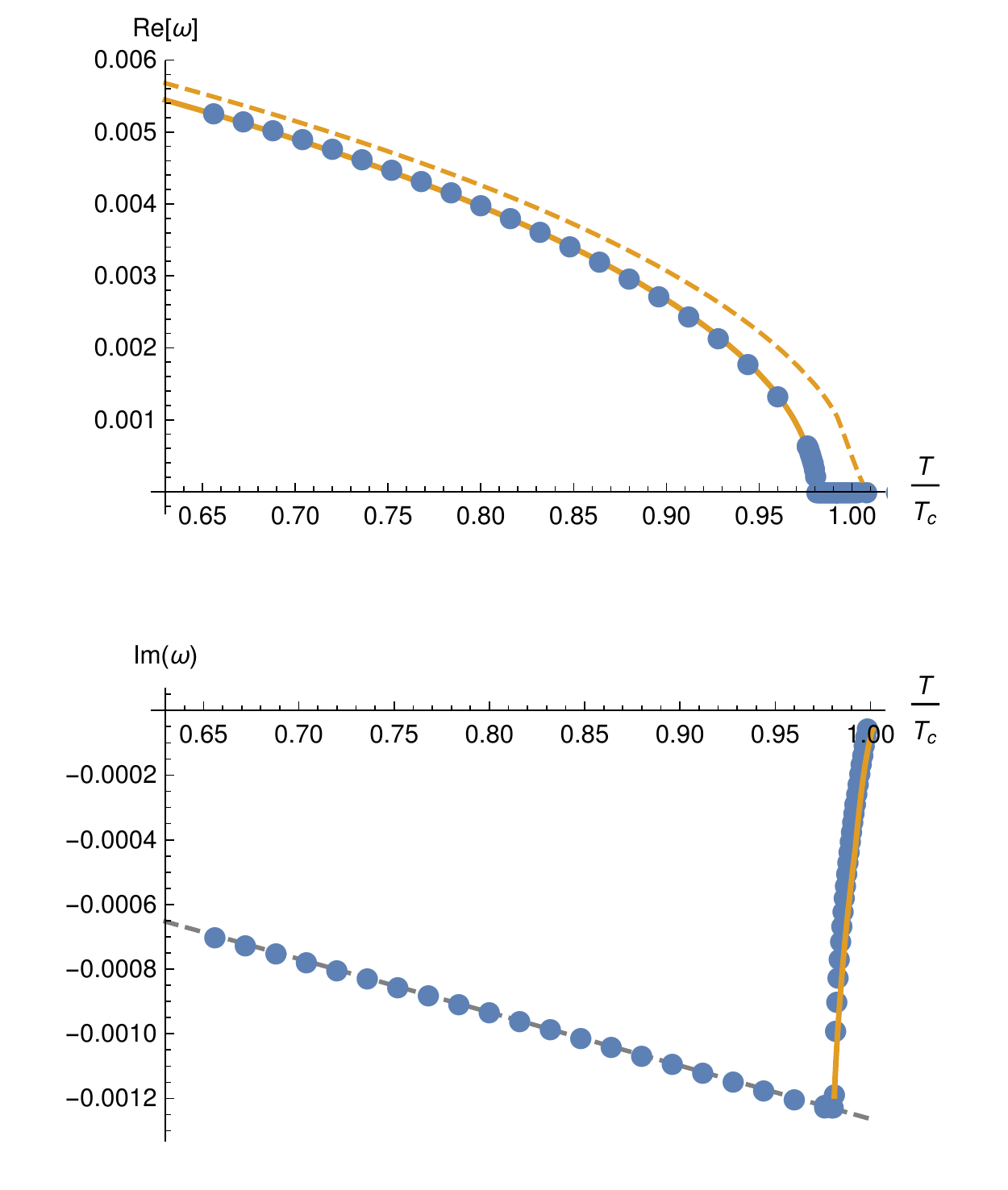}
\end{minipage}
\caption{\label{Fig:VaryT_QNMs} QNMs in the states with fixed explicit scale $\lambda=0.5$ and increasing temperature. As the temperature approaches the critical point $T_c$ the phonon QNMs become purely diffusive. Yellow curves show the fit using the order parameter scale \eqref{Jfit}, yellow dashed line displays the appropriately normalized order parameter $\langle J_2 \rangle$ itself. Left panel: trajectories of QNMs in the complex plane, Right panel: dependence of the real and imaginary parts on $T$. }
\end{figure}

Having studied the dependence of the phonon pinning frequency on the explicit symmetry breaking scale, let us turn to the study of its relation to the spontaneous symmetry breaking scale, which is set in our context by the order parameter corresponding to the vacuum expectation value of the helical current $\la J_2 \ra$. We can accomplish this by changing temperature, since the order parameter is clearly temperature dependent. In Fig.\,\ref{Fig:VaryT_QNMs} we show the evolution of the QNMs when we rise the temperature all the way to the critical point, 
while keeping the explicit scale $\lambda=0.5$ fixed.  

\begin{figure}[ht]
\center
\includegraphics[width=0.49\linewidth]{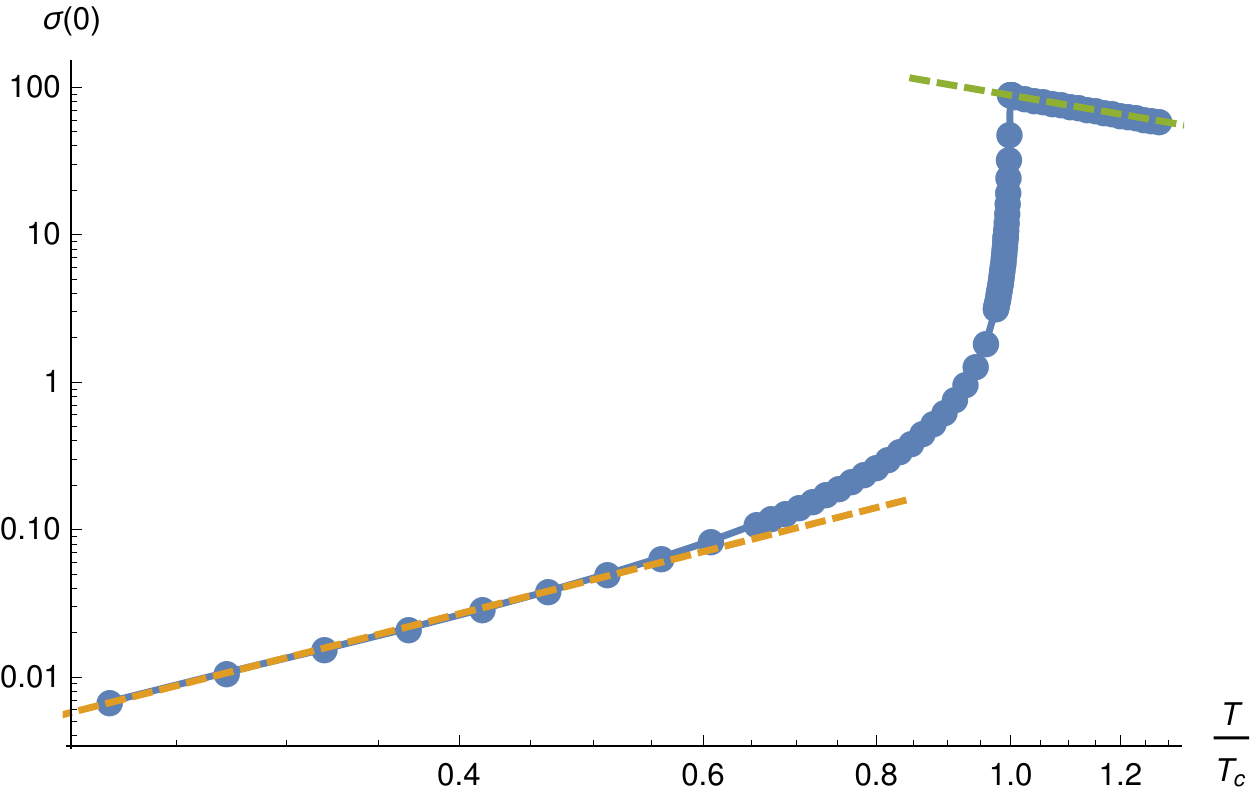}
\includegraphics[width=0.49\linewidth]{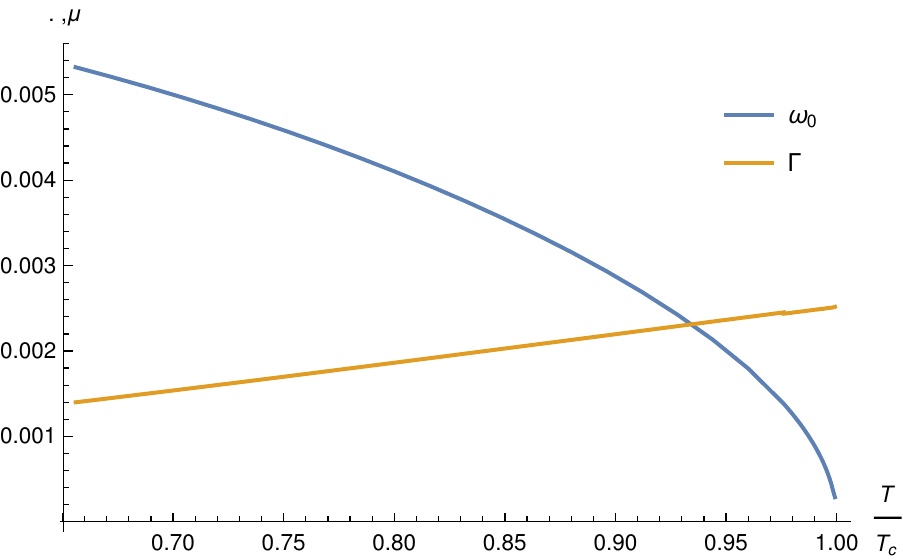}
\caption{\label{Fig:TSigmaLog} \textbf{Left panel:} DC conductivity scaling at low and high temperatures across the phase transition. \textbf{Right panel:} Relation between pinning frequency ($\omega_0$) and relaxation rate $\Gamma$. Clearly, the pseudo-Goldstone approximation ($\Gamma \ll \omega_0$) is violated near $T_c$}
\end{figure}

The striking feature of these spectra is that the phonon does not acquire a finite gap in the spectrum immediately after the phase transition. Instead it becomes an over-damped diffusive mode until the temperature is lowered further (down to $T\sim 0.98 T_c$ in the present case). 
The reason behind this is that exactly below the critical temperature the spontaneous symmetry breaking scale is infinitesimally small, while the explicit scale is fixed to a finite value. In this case the pinning frequency $\omega_0$ may happen to be smaller then the damping factor $\Gamma$ (see Fig.\ref{Fig:TSigmaLog}, right panel) and the poles in \eqref{cond}, which are
\begin{align}
\label{Jfit}
\omega_{1,2} = \frac{1}{2} \left(- i \Gamma \pm \sqrt{4 \omega_0^2 - \Gamma^2} \right), 
\end{align}
do indeed become purely imaginary. On the right panel of Fig.\,\ref{Fig:VaryT_QNMs} the yellow line shows the fit which we obtain from \eqref{Jfit} by assuming that the pinning frequency is proportional to the order parameter, while the damping is constant
\begin{equation}
\label{equ:omega_J_scaling}
\omega_0 \sim  \la J_2 \ra^1, \qquad \Gamma \sim \la J_2 \ra^0.
\end{equation}
The fit displays very good agreement in both real and imaginary parts, given that only one proportionality constant is fitted and we evaluate $\la J_2 \ra$ independently by analysing the background solutions. It would be interesting to analyze our results regarding the relaxation rate $\Gamma$ using the memory matrix techniques as in \cite{Hartnoll:2012rj,Lucas:2015lna,Lucas:2015vna,Delacretaz:2017zxd}.

\begin{figure}[t]
\center
\begin{minipage}{0.5 \linewidth}
\includegraphics[width=1.\linewidth]{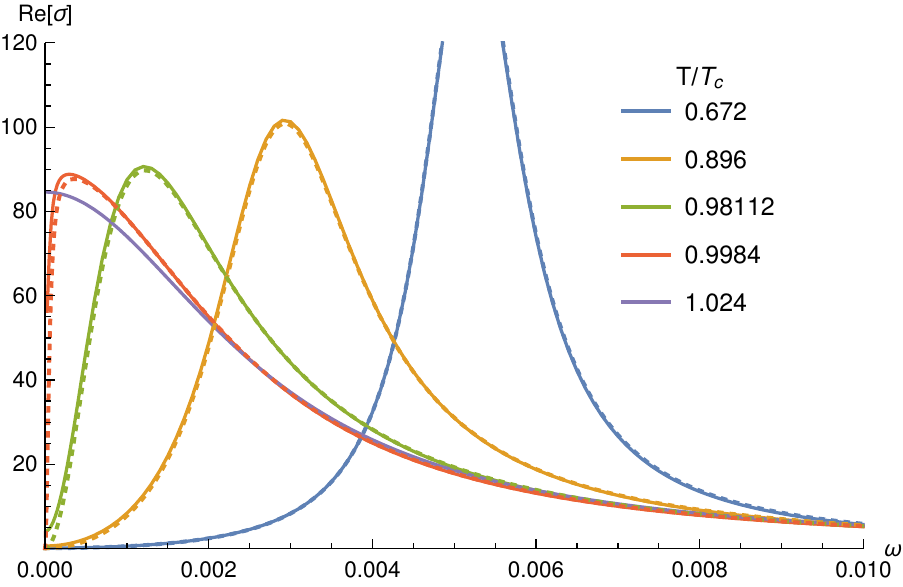}
\end{minipage}{}
\begin{minipage}{0.49 \linewidth}
\includegraphics[width=1.\linewidth]{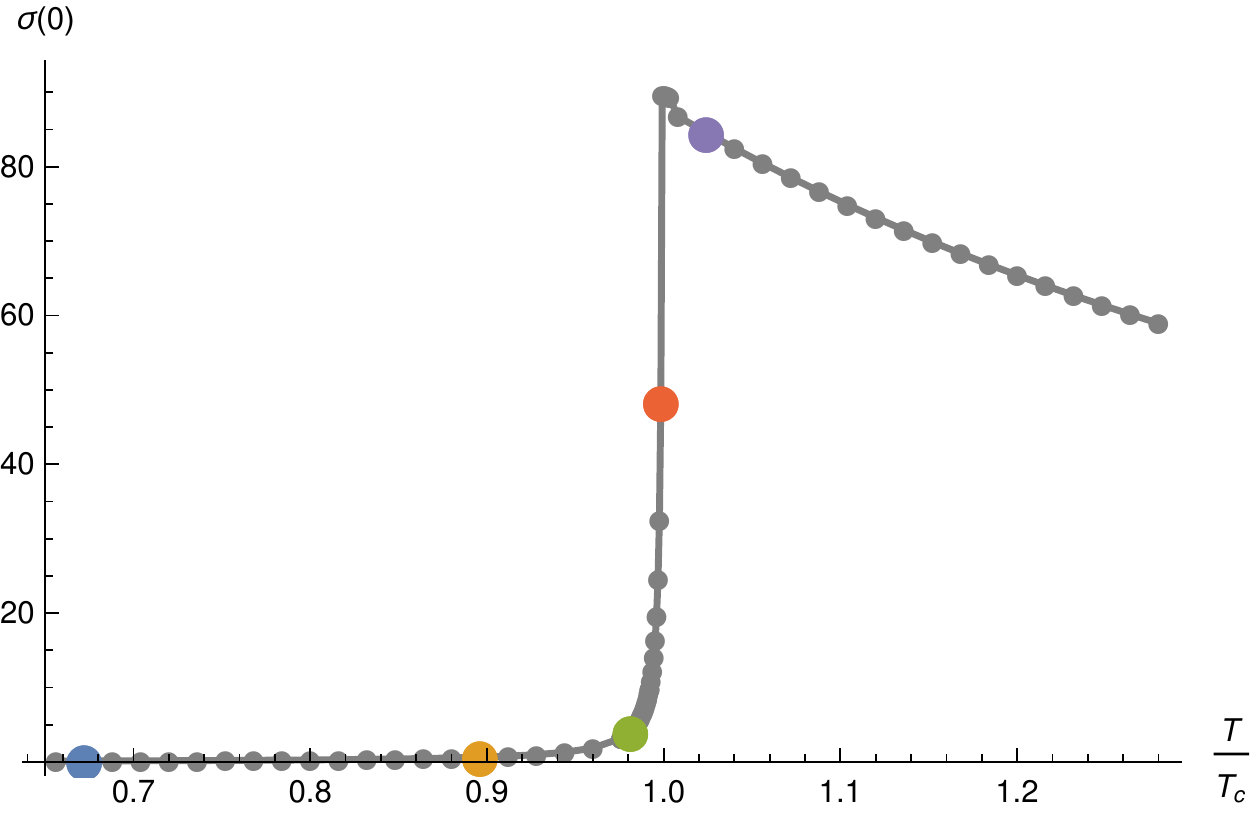}
\end{minipage}
\caption{\label{Fig:TAC} AC conductivity (Left panel) and its limit at zero frequency (Right panel) in the states with fixed explicit symmetry breaking ($\lambda=0.5$) and changing temperature. The dashed lines show the prediction of \eqref{cond} with parameters extracted from QNM spectrum. The solid lines are the numerical results. Color dots on the right panel mark the parameters for which the AC curves has been produced.}
\end{figure}

Finally, we investigate the AC and DC conductivity as a function of temperature. Fig.\,\ref{Fig:TAC} shows the AC conductivity fitted by the expression \eqref{cond} in both propagating and over-damped regimes. Again the fits are almost perfect for the whole range of temperatures. As it is seen from \eqref{cond} and \eqref{Jfit}, the position of the peak in AC conductivity is always set by $\omega_0$, which is positive even in the over-damped regime.
This good agreement is actually surprising given that in the regime $\Gamma > \omega_0$ the scale of explicit symmetry breaking gets larger then the scale of spontaneous symmetry breaking, which is formally beyond the region of validity of the formula \eqref{cond}. A slight mismatch can be discerned though in the small $\omega$ region.

The DC conductivity plot looks somewhat puzzling in the region near $T_c$. On one hand, given that the formula \eqref{cond} provides a reliable fit to the AC conductivity, one would expect, similarly to Fig.\,\ref{Fig:LamAC}, that DC conductivity is not affected by the contribution from the peak and 
stays constant all the way to the critical temperature. On the other hand, the DC plot clearly shows the hyperbolic-like upturn right below $T_c$. As  pointed out above, the explanation is that right next to the critical temperature the explicit symmetry breaking scale so strongly dominates over the spontaneous one (see right panel of Fig.\ref{Fig:TSigmaLog}), that one has to take into account the corrections to the formula \eqref{cond}, which is obtained in the gradient expansion approximation and as a deviation from the pure Goldstone mode. We discuss the possible corrections in the Appendix \ref{app-hydro}. 

Before closing this section let us comment on the ``metal-insulator'' transition happening due to the presence 
of the combined effect of the spontaneous and explicit breaking. 
It is best seen on the Log-Log plot of the DC conductivity, left panel of Fig.\,\ref{Fig:TSigmaLog}. Here we see that the high temperature metallic phase has conductivity which scales as
\begin{equation}
\sigma_{met} \sim T^{-1.6},
\end{equation}
while the low temperature state has the opposite scaling
\begin{equation}
\sigma_{ins} \sim T^{2.4}.
\end{equation}
As anticipated, at low temperature the system turns into the algebraic insulator with a power law fall of the conductivity, confirming 
that the metal-insulator transition is indeed induced by the pinning of a Goldstone mode. \\
The metallic/insulating nature of the system at low temperature, and the corresponding scalings of the conductivity, could be understood in terms of the ground states of the Bianchi VII spacetime considered. Previous studies of the IR fixed points in this direction have been performed in \cite{Donos:2012js,Donos:2014oha,Donos:2012gg,Donos:2017mhp}. It would be very interesting and useful to elucidate this point better in the future to have better control on the IR geometries and their transport properties.

\section{Discussion and Summary}
\label{sec:discussion}

We have explicitly obtained the dispersion relation of a 
Goldstone mode associated to the spontaneous breaking of translations within the holographic framework.  
This Goldstone can be identified as the longitudinal phonon of the corresponding ``electronic crystal". We also observed that the speed of sound mediated by this phonon is affected by the elasticity of the crystal.
After identifying
this mode in the spectrum of QNMs, we have introduced a source which explicitly breaks translational 
invariance and followed the trajectory of the Goldstone in the complex plane. 
We observe that the initially massless mode, i.e. the proper Goldstone boson, acquires a mass gap and a finite damping $\Gamma$  according to \eqref{pseudo}. 
We demonstrate that the pinning frequency $\omega_0$ is proportional to both the explicit breaking amplitude \eqref{equ:omega_lambda_scaling} and the spontaneous breaking order parameter \eqref{equ:omega_J_scaling} in accordance with the general expectations for the pseudo-Goldstone mode
\begin{equation}
\omega_0 \sim \lambda \la J_2 \ra.
\end{equation}
%
The pinning frequency is a direct manifestation of the pinning mechanism due to the interplay between two lattice structures while the damping parameter $\Gamma$, which is only related to the explicit scale $\lambda$ is just a reflection of the a finite momentum relaxation rate $\tau_{rel}$.
Both parameters become zero in the limit of purely spontaneous breaking $\lambda=0$ and one recovers the usual massless phonon.
In other words, our model provides a concrete realization of the interplay of spontaneous and explicit breaking of translations in perfect 
agreement with the effective field theory expectations \cite{Delacretaz:2017zxd}.

Furthermore, we have computed the 
optical conductivity, which offers a complementary view of the pinning physics. More concretely, 
tracing the behaviour of a purely metallic solution as we decrease the temperature below $T_c$, 
we observe the shift to higher frequencies of the zero frequency Drude peak present in the purely 
explicit breaking scenario. 
Due to this frequency shift, there is a drastic drop in the DC conductivity across the critical temperature, 
which occurs in such a way that our initially metallic solutions become algebraic insulators, i.e. 
$\sigma_{DC}$ decreases with temperature as a power law at low temperatures. 
Moreover, the value of $\sigma_{DC}$ is nearly independent of the symmetry breaking parameter in this insulating state, in agreement with the hydrodynamic expectations which identifies $\sigma_{DC}$ in the presence of pinning to the so called incoherent conductivity $\sigma_Q$ \cite{Davison:2015taa,Delacretaz:2017zxd}.

Our work opens up a new direction towards the study of phonons and elasticity within the holographic picture. Along these lines, 
it would be interesting to consider other physical observables which are naturally strongly influenced by the phononic 
contribution, such as the heat capacity and the thermal conductivity. In order to give these studies a firmer basis, it would be 
useful to build up a precise framework to compute the elastic moduli and the viscosities via Kubo formulae in presence of explicit and spontaneous breaking of translations, using for example a gradient expansion. The study of the elastic properties of the model we considered is certainly an exciting direction to pursue. The related questions are many: from a precise definition of the elastic modulus $\kappa$ and a formal derivation of the formula for the phonons speed \eqref{speeds} to a direct comparison with the numerical results to assess the agreement with the theoretical expectations.

Moreover, it would be interesting to further investigate the transition region, when the DC conductivity starts rising when approaching $T_c$ from below, since this behaviour is naively forbidden by the formula \eqref{cond}. The possible breakdown of this approximation may already 
happen in the regime where the spontaneous order parameter is infinitesimally small, which allows for a perturbative access to this region 
in parameter space. 

We have obtained our results relying on the technical simplifications that arise from considering a homogeneous model 
with Bianchi VII helical symmetry. The excellent matching with the results obtained using a rather generic gradient expansion 
\eqref{pseudo}, \eqref{cond} is indeed a posteriori justification of the validity of this model. However, the implementation 
of the ideas exposed here in a truly inhomogeneous setup would allow us to uncover interesting physics, such as the 
the propagation of a transverse phonon and the consequences of the lock in between the spontaneous and explicit structure.

We hope to return to these and other related issues in the near future.

 \acknowledgments

We thank Andrea Amoretti, Aron Beekman, Blaise Gouteraux, Matti Jarvinen, Matthew Lippert, Rene Meyer, Sang-Jin Sin, Koenraad Schalm and Jan Zaanen  for useful discussions and comments about this work and the topics considered.
We are particularly grateful to Matthew Lippert, Niko Jokela, Marti Jarvinen, L.Alberte, M.Ammon, O.Pujolas and A.Jimenez for sharing with us informations about their unpublished works.
We thank organizers of the workshop ``Disorder in Condensed Matter and Black Holes'' held in Lorentz Center, Leiden University, in January 2017 where this project has been initiated.
The work of T.A. was supported by the European Research Council under the European Union's Seventh Framework Programme 
(ERC Grant agreement 307955). He also acknowledges the partial support of the Newton-Picarte Grant 20140053. 
The work of A.K. is supported in part by the VICI grant of Koenraad Schalm from the Netherlands Organization for Scientific Research (NWO), by the Netherlands Organization for Scientific Research/Ministry of Science and Education (NWO/OCW) and by the Foundation for Research into Fundamental Matter (FOM). A.K. is partially supported by RFBR grant 15-02-02092a. 
M.B. is supported in part by the Advanced ERC grant SM-grav, No 669288. M.B. would like to thank the Asia Pacific Center for Theoretical Physics, Hanyang University and Gwangju Institute of Science and Technology for the hospitality during the completion of this work. The work of N.P. is supported by DPST scholarship from the Thai government and by Leiden University.

\appendix

\section{Low energy spectrum of a theory without translational symmetry}\label{app-hydro}

In this appendix, we summarize the spectrum of the effective theory where translational symmetry breaking occurs in one direction. 
We first deal with the scenario where the symmetry is only spontaneously broken, and then include momentum relaxation. 
Our analysis closely follows the framework of \cite{Delacretaz:2017zxd}, where the effective hydrodynamic theory for charge density wave (CDW) states has been constructed. We refer interested readers to \cite{Martin:1972,Zippelius:1980,Delacretaz:2017zxd} for more details regarding the setup and also \cite{Nicolis:2013lma,Delacretaz:2014jka,Nicolis:2015sra} for a modern discussion from the effective action point of view.

We restrict ourselves to the scenario where the wave number $q^i$ is parallel to the direction of symmetry breaking, described by a unit vector $n^i$. We also assume that the system is isotropic in the plane perpendicular to $n^i$. Using the Laplace transform method \cite{Kadanoff} (see also \cite{Hartnoll:2007ih,Kovtun:2012rj} for a modern review), one is able to find the spectrum of this system. It turns out that the spectrum contains  two sound poles and three diffusive poles which can be written schematically as 
\begin{equation}
\omega_\text{S} = c_s q -\frac{i}{2} \mathcal{D}_s q^2 ,\quad \omega_\text{T} = -i\left(\frac{\eta_\perp}{\varepsilon+p}\right)q^2,\quad \omega_{1,2} = -i \mathcal{D}_{1,2}\,q^2
\end{equation}
where the speed of the propagating mode $\omega_S$ is given by \eqref{speeds}. The mode $\omega_T$ is the diffusive mode of the transverse momentum, with $\eta_\perp$ being one of the shear viscosities in an anisotropic system. The diffusion constants $\mathcal{D}_{1,2}$ are complicated functions of the thermodynamic quantities, transport coefficients and elastic modulus. However, in the limit of small elastic modulus, we find that 
\begin{equation}
 \mathcal{D}_1 = \mathcal{D}_\rho  + \kappa \, \delta\mathcal{D}_1 + \mathcal{O}(\kappa^2),\qquad \mathcal{D}_2 = \kappa \,\delta\mathcal{D}_2 + \mathcal{O}(\kappa^2)
\end{equation}
where $\delta\mathcal{D}_{1,2}$ are functions of transport coefficients and thermodynamics quantities. Here $\mathcal{D}_\rho$ is the thermoelectric diffusion in a charged relativistic fluid (see e.g. \cite{Kovtun:2012rj,Davison:2015taa}) and $\kappa$ the elastic modulus introduced in \eqref{speeds}. 
We can therefore deduce that the mode of frequency $\omega_1$ is the thermoelectric diffusion with the diffusion constant modified by the elastic modulus. On the other hand, the diffusive mode $\omega_2$ is unique to the spontaneous translational symmetry broken phase as its diffusion constant vanishes when the elastic modulus goes to zero. 
Following \cite{Delacretaz:2017zxd}, we refer to $\omega_2$ as a crystal diffusive mode\footnote{In \cite{Delacretaz:2017zxd}, the results for the system with spontaneous symmetry breaking in one dimension are presented for an arbitrary propagation angle $\cos\theta \sim n_i q^i$. In such scenario, one finds the longitudinal sound, the thermoelectric diffusion together with an additional propagating mode: the transverse sound. However, a careful analysis of the Green's function indicates that the limits $q\to 0$ and $\theta \to 0$ do not commute. The result presented here is obtained by first setting $\theta \to 0$ and then perform the small gradient expansion in $q$. Following this procedure, one notices 
that the two transverse sound poles collide with each other on the imaginary axis and turn into the transverse diffusive mode $\omega_T$ and the crystal diffusive mode $\omega_1$ as one decreases $\theta$ from $\theta = \pi/2$ toward $\theta = 0$. Similar behaviour is also found in the study of magnetosonic waves in dissipative relativistic plasmas \cite{Hernandez:2017mch,Grozdanov:2017kyl}.}. These results are summarized in Table \ref{table:spectrumPureSpontaneous} 
\begin{table}[tbh]
\begin{center}
    \begin{tabular}{ | l | l |}
    \hline
    Modes & Dispersion Relation (small $\kappa$) \\ \hline
    Sound (S) & $\omega_\text{S} = c_s q -\frac{i}{2} \mathcal{D}_s q^2$    \\ \hline
    Transverse Diffusive (T) & $\omega_\text{T} = -i\left(\frac{\eta_\perp}{\varepsilon+p}\right)q^2$   \\ \hline
    Thermoelectric diffusion (D1) & $\omega_1 =-i\left( \mathcal{D}_\rho+ \kappa \,\delta \mathcal{D}_1 \right) q^2$  \\
    \hline
    Crystal diffusion (D2) & $\omega_2 =-i \kappa \,\delta \mathcal{D}_2\,  q^2$    \\
    \hline
    \end{tabular}
    \caption{A table summarizing the low energy modes in the spontaneous symmetry broken phase. The explicit expression for $c_0, \mathcal{D}_\rho$ can be found in the in \cite{Kovtun:2012rj}. The expression for $\delta\mathcal{D}_2$ can be obtain straightforwardly from the setup in the appendix A of \cite{Delacretaz:2017zxd} taking the limit $\theta = 0$ first, then Taylor expand $q$ around $q=0$. These modes are observed in our holographic model as illustrated in Fig.\,\ref{Fig:VaryQ_QNMs}.}
\label{table:spectrumPureSpontaneous}
\end{center}
\end{table}

The explicit translational symmetry breaking can be introduced by adding momentum relaxation term to the r.h.s. of 
the conservation of momentum equation. Moreover, as argued in e.g. \cite{Delacretaz:2017zxd}, the explicit translational 
symmetry breaking will also generate a gap $\omega_0$ for the Goldstone boson $\phi$. As a result, the conservation of momentum is modified as 
\begin{equation}\label{modConservationofMomentum}
\partial_t \pi^i + \partial_j \tau^{ij} = 0,\quad \Rightarrow\quad \partial_t \pi^i + \partial_j \tau^{ij} = -\Gamma^{ij}\pi_j -\chi_{\pi\pi}\,\omega_0^2\, n^i \phi
\end{equation}
where $\pi^i$ is the momentum and $\tau^{ij}$ is the stress tensor. If one assumes that the constitutive relation is independent of $\Gamma^{ij}$ and $\omega_0$, one finds that the pole of spatially resolved AC conductivity $\sigma(\omega,k)$ is governed by the pseudo-Goldstone dispersion relation \eqref{pseudo} and the AC conductivity can be expressed as in \eqref{cond} where $\Gamma$ is the longitudinal part of $\Gamma^{ij}$. 
Note however that, as we neglect the contribution from these momentum relaxation terms in the constitutive relation, we are unable to reliably describe the dissipative effect in this scenario\footnote{For a hydrodynamic-like framework where the momentum relaxation is taken into account consistently see e.g. \cite{Blake:2015epa,Blake:2015hxa,Burikham:2016roo} where the symmetry is broken explicitly by spatial dependent scalar fields, by chemical potential \cite{Lucas:2015sya} and by curved background metric \cite{Banks:2016krz,Scopelliti:2017sga}.  }. Nevertheless, the effective theory of \eqref{modConservationofMomentum}, is still able to capture essential properties of various quantities when the momentum relaxation rate is small. On the other hand, when the momentum relaxation rate is large, it is clear that the spectrum will deviate from the formula in \eqref{pseudo}. Furthermore, one can deduce that the propagating mode will disappear and the transport phenomena will be fully governed by diffusion \cite{Hartnoll:2014lpa,Davison:2014lua,Delacretaz:2017zxd}. This phenomenon where the sound mode turns into diffusive modes was first discussed in the model of \cite{Andrade:2013gsa} in \cite{Davison:2014lua}.   

Lastly, we emphasize that the model considered in this appendix (as well as in \cite{Delacretaz:2017zxd}) is  isotropic in the plane perpendicular to $n^i$ while the Bianchi VII setup is not. As a result, the detailed structures (such as elastic moduli and independent transport coefficients) of the effective theory in this section and our holographic model are expected to be somewhat different. 

Nevertheless, this hydrodynamic model shares many qualitative features with the low energy mode in the Bianchi VII setup, 
due to the fact that we only consider the propagation longitudinal to $n^i$. Because of this, the structure of the lattice/crystal perpendicular 
to $n^i$ should not change the obtained qualitative behaviour of the low energy modes.

\section{Linearized modes, asymptotics and De Donder gauge}\label{app-linear}
\label{app:linear_eqs}

The system of equations which governs our linear modes corresponds to 25 coupled ODEs for 
{\it all} the components of the metric and vectors fields. 
We remove the gauge redundancy by imposing the DeDonder and Lorentz conditions
\begin{align}
\label{DeDonder}
	\tau^g_\mu &:= \nabla^{\mu}( \delta g_{\mu \nu} - \frac{1}{2} g_{\mu \nu} \delta g ) = 0  \\
 	\tau^A &:= \nabla^\mu \delta A_\mu = 0 \\
	 \tau^B &:= \nabla^\mu \delta B_\mu = 0
\end{align}
Following \cite{Rangamani:2015hka}, we implement this by adding the terms $\nabla_{( \mu} \tau^g_{\nu)}$, $\nabla_\mu \tau^A$,
$\nabla_\mu \tau^B$ to the Einstein and Maxwell equations, respectively, with the appropriate coefficients. These coefficients are fixed 
by demanding that the resulting equations of motion are of Hyperbolic form. 

We demand the fluctuations to satisfy ingoing boundary conditions at the horizon, which is best seen by expressing the metric 
and gauge field in terms of Eddington-Finkelstein coordinates. In practice, this amounts to extracting a factor $(1- u)^{\beta_n}$, in 
all the modes, where
\begin{equation}
	\beta_n = - i \frac{\omega}{4 \pi T} - n
\end{equation}
Here, $n = 0,1,2$, is determined by the number of indices in the $u$-direction in the profile, e.g. $n=2$ for $\delta g_{uu}$. The 
near horizon equations then require a set of relations among the leading and subleading terms in the regular expansions for the fields. 
Some of these are to be imposed as boundary conditions, while the others can be checked a posteriori by consistency of the equations of motion, 
see \cite{Rangamani:2015hka} for a more detailed discussion. 

The near boundary expansions of the linear perturbations have the same form as the non-linear counterparts, and can 
be found in \cite{Erdmenger:2015qqa}.

\section{Numerical linear analysis: QNMs, AC conductivity and $\sigma(0)$} \label{app-QNM}
Once the linearized equations of motion, including boundary conditions, for the excitations are discretized using the pseudospectral method, they can be expressed as a linear system of equations on the values of all coupled modes at every grid point. Denoting the matrix operator of the corresponding homogeneous system as $M$ we can write down the equations for AC conductivity as 
\begin{equation}
\label{AC_linear}
M \cdot \vec{v} = (\dots,A_x^{(0)},\dots,\dots)^{T} \equiv \vec{A},
\end{equation}
where the position of the single entry in the right hand side corresponds to the boundary condition equation for $A_x$ field. Hence the profile of the solution with the source is simply
\begin{equation}
\vec{v} = M^{-1} \cdot \vec{A},
\end{equation}
The responce is read off from the grid values encoded in $v$ vector using \eqref{sigma_definition}.

There is a neat way to evaluate directly the zero frequency limit of AC conductivity. Given that the linearized equations of motion are second order in $\omega$ we can represent the matrix $M$ exaclty as
\begin{equation}
\label{M_expansion}
M = M_0 + i \omega M_1 + \omega^2 M_2
\end{equation}
At small $\omega$ the solution to \eqref{AC_linear} can be expanded as $\vec{v} = \vec{v}_0 + i \omega \vec{v}_1 + \dots$. Expanding \eqref{AC_linear} in first two orders we get
\begin{align}
M_0 \cdot \vec{v}_0 &= \vec{A}, \\
M_1 \cdot \vec{v}_0 + M_0 \cdot \vec{v}_1 &= 0.
\end{align}
For momentum $q=0$, which is the case for AC conductivity calculation, both equations are manifestly real and we get the solution
\begin{align}
\vec{v} &= \vec{v}_0 + i \omega \vec{v}_1 \\
\vec{v}_0 &= M_0^{-1} \cdot \vec{A} \\ 
\vec{v}_1 &= - M_0^{-1} \cdot M_1 \cdot \vec{v}_0
\end{align}
According to \eqref{sigma_definition}, the real part of the conductivity is related to the subleading coefficient in the imaginary part of the solution devided by frequency, hence one can read off the zero frequency limit of AC conductivity as
\begin{equation}
\sigma(0) \equiv \lim_{\omega \rar 0} \sigma (\omega) \sim D_z^2 \cdot \vec{v}_1,
\end{equation}
where $D_z^2$ is a discretized version of the second derivative.

In order to study quasinormal modes one has to analyse the eigenvalues of the matrix~$M$. Given the form \eqref{M_expansion}, this task boils down to the well known quadratic eigenvalue problem, which we solve using \textit{Wolfram Mathematica} \texttt{Eigenvalue} routine \cite{Mathematica10}.

\label{app-num}

\bibliographystyle{JHEP-2}
\bibliography{inhom_stripes_lattice}

\providecommand{\href}[2]{#2}\begingroup\raggedright\begin{thebibliography}{10}

\bibitem{Hartnoll:2012rj}
S.~A. Hartnoll and D.~M. Hofman, {\it {Locally Critical Resistivities from
  Umklapp Scattering}},  {\em Phys. Rev. Lett.} {\bf 108} (2012) 241601
  [\href{http://arXiv.org/abs/1201.3917}{{\tt 1201.3917}}].

\bibitem{Lucas:2015vna}
A.~Lucas, {\it {Conductivity of a strange metal: from holography to memory
  functions}},  {\em JHEP} {\bf 03} (2015) 071
  [\href{http://arXiv.org/abs/1501.05656}{{\tt 1501.05656}}].

\bibitem{Nitta:2017mgk}
M.~Nitta, S.~Sasaki and R.~Yokokura, {\it {Spatially Modulated Vacua in
  Relativistic Field Theories}},  \href{http://arXiv.org/abs/1706.02938}{{\tt
  1706.02938}}.

\bibitem{Delacretaz:2016ivq}
L.~V. Delacrétaz, B.~Goutéraux, S.~A. Hartnoll and A.~Karlsson, {\it {Bad
  Metals from Fluctuating Density Waves}},
  \href{http://arXiv.org/abs/1612.04381}{{\tt 1612.04381}}.

\bibitem{Delacretaz:2017zxd}
L.~V. Delacrétaz, B.~Goutéraux, S.~A. Hartnoll and A.~Karlsson, {\it
  {Hydrodynamic transport in fluctuating charge density waves}},
  \href{http://arXiv.org/abs/1702.05104}{{\tt 1702.05104}}.

\bibitem{RevModPhys.60.1129}
G.~Gr\"uner, {\it The dynamics of charge-density waves},  {\em Rev. Mod. Phys.}
  {\bf 60} (Oct, 1988) 1129--1181.

\bibitem{Zaanen:2015oix}
J.~Zaanen, Y.-W. Sun, Y.~Liu and K.~Schalm, {\em {Holographic Duality in
  Condensed Matter Physics}}.
\newblock Cambridge Univ. Press, 2015.

\bibitem{Nakamura:2009tf}
S.~Nakamura, H.~Ooguri and C.-S. Park, {\it {Gravity Dual of Spatially
  Modulated Phase}},  {\em Phys. Rev.} {\bf D81} (2010) 044018
  [\href{http://arXiv.org/abs/0911.0679}{{\tt 0911.0679}}].

\bibitem{Ooguri:2010kt}
H.~Ooguri and C.-S. Park, {\it {Holographic End-Point of Spatially Modulated
  Phase Transition}},  {\em Phys. Rev.} {\bf D82} (2010) 126001
  [\href{http://arXiv.org/abs/1007.3737}{{\tt 1007.3737}}].

\bibitem{Donos:2012wi}
A.~Donos and J.~P. Gauntlett, {\it {Black holes dual to helical current
  phases}},  {\em Phys. Rev.} {\bf D86} (2012) 064010
  [\href{http://arXiv.org/abs/1204.1734}{{\tt 1204.1734}}].

\bibitem{Donos:2013gda}
A.~Donos and J.~P. Gauntlett, {\it {Holographic charge density waves}},  {\em
  Phys. Rev.} {\bf D87} (2013), no.~12 126008
  [\href{http://arXiv.org/abs/1303.4398}{{\tt 1303.4398}}].

\bibitem{Donos:2011ff}
A.~Donos and J.~P. Gauntlett, {\it {Holographic helical superconductors}},
  {\em JHEP} {\bf 12} (2011) 091 [\href{http://arXiv.org/abs/1109.3866}{{\tt
  1109.3866}}].

\bibitem{Donos:2011bh}
A.~Donos and J.~P. Gauntlett, {\it {Holographic striped phases}},  {\em JHEP}
  {\bf 08} (2011) 140 [\href{http://arXiv.org/abs/1106.2004}{{\tt 1106.2004}}].

\bibitem{Donos:2011pn}
A.~Donos, J.~P. Gauntlett and C.~Pantelidou, {\it {Magnetic and Electric AdS
  Solutions in String- and M-Theory}},  {\em Class. Quant. Grav.} {\bf 29}
  (2012) 194006 [\href{http://arXiv.org/abs/1112.4195}{{\tt 1112.4195}}].

\bibitem{Donos:2011qt}
A.~Donos, J.~P. Gauntlett and C.~Pantelidou, {\it {Spatially modulated
  instabilities of magnetic black branes}},  {\em JHEP} {\bf 01} (2012) 061
  [\href{http://arXiv.org/abs/1109.0471}{{\tt 1109.0471}}].

\bibitem{Rozali:2012es}
M.~Rozali, D.~Smyth, E.~Sorkin and J.~B. Stang, {\it {Holographic Stripes}},
  {\em Phys. Rev. Lett.} {\bf 110} (2013), no.~20 201603
  [\href{http://arXiv.org/abs/1211.5600}{{\tt 1211.5600}}].

\bibitem{Donos:2013woa}
A.~Donos, J.~P. Gauntlett and C.~Pantelidou, {\it {Competing p-wave orders}},
  {\em Class. Quant. Grav.} {\bf 31} (2014) 055007
  [\href{http://arXiv.org/abs/1310.5741}{{\tt 1310.5741}}].

\bibitem{Withers:2013kva}
B.~Withers, {\it {The moduli space of striped black branes}},
  \href{http://arXiv.org/abs/1304.2011}{{\tt 1304.2011}}.

\bibitem{Withers:2013loa}
B.~Withers, {\it {Black branes dual to striped phases}},  {\em Class. Quant.
  Grav.} {\bf 30} (2013) 155025 [\href{http://arXiv.org/abs/1304.0129}{{\tt
  1304.0129}}].

\bibitem{Jokela:2014dba}
N.~Jokela, M.~Jarvinen and M.~Lippert, {\it {Gravity dual of spin and charge
  density waves}},  {\em JHEP} {\bf 12} (2014) 083
  [\href{http://arXiv.org/abs/1408.1397}{{\tt 1408.1397}}].

\bibitem{Withers:2014sja}
B.~Withers, {\it {Holographic Checkerboards}},  {\em JHEP} {\bf 09} (2014) 102
  [\href{http://arXiv.org/abs/1407.1085}{{\tt 1407.1085}}].

\bibitem{Krikun:2015tga}
A.~Krikun, {\it {Phases of holographic d-wave superconductor}},  {\em JHEP}
  {\bf 10} (2015) 123 [\href{http://arXiv.org/abs/1506.05379}{{\tt
  1506.05379}}].

\bibitem{Erdmenger:2013zaa}
J.~Erdmenger, X.-H. Ge and D.-W. Pang, {\it {Striped phases in the holographic
  insulator/superconductor transition}},  {\em JHEP} {\bf 11} (2013) 027
  [\href{http://arXiv.org/abs/1307.4609}{{\tt 1307.4609}}].

\bibitem{Cremonini:2016rbd}
S.~Cremonini, L.~Li and J.~Ren, {\it {Holographic Pair and Charge Density
  Waves}},  \href{http://arXiv.org/abs/1612.04385}{{\tt 1612.04385}}.

\bibitem{Ling:2014saa}
Y.~Ling, C.~Niu, J.~Wu, Z.~Xian and H.-b. Zhang, {\it {Metal-insulator
  Transition by Holographic Charge Density Waves}},  {\em Phys. Rev. Lett.}
  {\bf 113} (2014) 091602 [\href{http://arXiv.org/abs/1404.0777}{{\tt
  1404.0777}}].

\bibitem{Cai:2017qdz}
R.-G. Cai, L.~Li, Y.-Q. Wang and J.~Zaanen, {\it {Intertwined order and
  holography: the case of the pair density wave}},
  \href{http://arXiv.org/abs/1706.01470}{{\tt 1706.01470}}.

\bibitem{Donos:2012js}
A.~Donos and S.~A. Hartnoll, {\it {Interaction-driven localization in
  holography}},  {\em Nature Phys.} {\bf 9} (2013) 649--655
  [\href{http://arXiv.org/abs/1212.2998}{{\tt 1212.2998}}].

\bibitem{Donos:2013eha}
A.~Donos and J.~P. Gauntlett, {\it {Holographic Q-lattices}},  {\em JHEP} {\bf
  04} (2014) 040 [\href{http://arXiv.org/abs/1311.3292}{{\tt 1311.3292}}].

\bibitem{Andrade:2013gsa}
T.~Andrade and B.~Withers, {\it {A simple holographic model of momentum
  relaxation}},  {\em JHEP} {\bf 05} (2014) 101
  [\href{http://arXiv.org/abs/1311.5157}{{\tt 1311.5157}}].

\bibitem{Donos:2014oha}
A.~Donos, B.~Goutéraux and E.~Kiritsis, {\it {Holographic Metals and
  Insulators with Helical Symmetry}},  {\em JHEP} {\bf 09} (2014) 038
  [\href{http://arXiv.org/abs/1406.6351}{{\tt 1406.6351}}].

\bibitem{Donos:2014uba}
A.~Donos and J.~P. Gauntlett, {\it {Novel metals and insulators from
  holography}},  {\em JHEP} {\bf 06} (2014) 007
  [\href{http://arXiv.org/abs/1401.5077}{{\tt 1401.5077}}].

\bibitem{Gouteraux:2014hca}
B.~Goutéraux, {\it {Charge transport in holography with momentum
  dissipation}},  {\em JHEP} {\bf 04} (2014) 181
  [\href{http://arXiv.org/abs/1401.5436}{{\tt 1401.5436}}].

\bibitem{Taylor:2014tka}
M.~Taylor and W.~Woodhead, {\it {Inhomogeneity simplified}},  {\em Eur. Phys.
  J.} {\bf C74} (2014), no.~12 3176 [\href{http://arXiv.org/abs/1406.4870}{{\tt
  1406.4870}}].

\bibitem{Horowitz:2012ky}
G.~T. Horowitz, J.~E. Santos and D.~Tong, {\it {Optical Conductivity with
  Holographic Lattices}},  {\em JHEP} {\bf 07} (2012) 168
  [\href{http://arXiv.org/abs/1204.0519}{{\tt 1204.0519}}].

\bibitem{Horowitz:2013jaa}
G.~T. Horowitz and J.~E. Santos, {\it {General Relativity and the Cuprates}},
  {\em JHEP} {\bf 06} (2013) 087 [\href{http://arXiv.org/abs/1302.6586}{{\tt
  1302.6586}}].

\bibitem{Donos:2014yya}
A.~Donos and J.~P. Gauntlett, {\it {The thermoelectric properties of
  inhomogeneous holographic lattices}},  {\em JHEP} {\bf 01} (2015) 035
  [\href{http://arXiv.org/abs/1409.6875}{{\tt 1409.6875}}].

\bibitem{Rangamani:2015hka}
M.~Rangamani, M.~Rozali and D.~Smyth, {\it {Spatial Modulation and
  Conductivities in Effective Holographic Theories}},  {\em JHEP} {\bf 07}
  (2015) 024 [\href{http://arXiv.org/abs/1505.05171}{{\tt 1505.05171}}].

\bibitem{Erdmenger:2015qqa}
J.~Erdmenger, B.~Herwerth, S.~Klug, R.~Meyer and K.~Schalm, {\it {S-Wave
  Superconductivity in Anisotropic Holographic Insulators}},  {\em JHEP} {\bf
  05} (2015) 094 [\href{http://arXiv.org/abs/1501.07615}{{\tt 1501.07615}}].

\bibitem{Andrade:2014xca}
T.~Andrade and S.~A. Gentle, {\it {Relaxed superconductors}},  {\em JHEP} {\bf
  06} (2015) 140 [\href{http://arXiv.org/abs/1412.6521}{{\tt 1412.6521}}].

\bibitem{Kim:2015dna}
K.-Y. Kim, K.~K. Kim and M.~Park, {\it {A Simple Holographic Superconductor
  with Momentum Relaxation}},  {\em JHEP} {\bf 04} (2015) 152
  [\href{http://arXiv.org/abs/1501.00446}{{\tt 1501.00446}}].

\bibitem{Ling:2014laa}
Y.~Ling, P.~Liu, C.~Niu, J.-P. Wu and Z.-Y. Xian, {\it {Holographic
  Superconductor on Q-lattice}},  {\em JHEP} {\bf 02} (2015) 059
  [\href{http://arXiv.org/abs/1410.6761}{{\tt 1410.6761}}].

\bibitem{Baggioli:2015zoa}
M.~Baggioli and M.~Goykhman, {\it {Phases of holographic superconductors with
  broken translational symmetry}},  {\em JHEP} {\bf 07} (2015) 035
  [\href{http://arXiv.org/abs/1504.05561}{{\tt 1504.05561}}].

\bibitem{Baggioli:2015dwa}
M.~Baggioli and M.~Goykhman, {\it {Under The Dome: Doped holographic
  superconductors with broken translational symmetry}},  {\em JHEP} {\bf 01}
  (2016) 011 [\href{http://arXiv.org/abs/1510.06363}{{\tt 1510.06363}}].

\bibitem{Andrade:2015iyf}
T.~Andrade and A.~Krikun, {\it {Commensurability effects in holographic
  homogeneous lattices}},  {\em JHEP} {\bf 05} (2016) 039
  [\href{http://arXiv.org/abs/1512.02465}{{\tt 1512.02465}}].

\bibitem{Andrade:2017leb}
T.~Andrade and A.~Krikun, {\it {Commensurate lock-in in holographic
  non-homogeneous lattices}},  {\em JHEP} {\bf 03} (2017) 168
  [\href{http://arXiv.org/abs/1701.04625}{{\tt 1701.04625}}].

\bibitem{Jokela:2016xuy}
N.~Jokela, M.~Jarvinen and M.~Lippert, {\it {Holographic sliding stripes}},
  \href{http://arXiv.org/abs/1612.07323}{{\tt 1612.07323}}.

\bibitem{Cremonini:2017usb}
S.~Cremonini, L.~Li and J.~Ren, {\it {Intertwined Orders in Holography: Pair
  and Charge Density Waves}},  \href{http://arXiv.org/abs/1705.05390}{{\tt
  1705.05390}}.

\bibitem{Vegh:2013sk}
D.~Vegh, {\it {Holography without translational symmetry}},
  \href{http://arXiv.org/abs/1301.0537}{{\tt 1301.0537}}.

\bibitem{Blake:2013owa}
M.~Blake, D.~Tong and D.~Vegh, {\it {Holographic Lattices Give the Graviton an
  Effective Mass}},  {\em Phys. Rev. Lett.} {\bf 112} (2014), no.~7 071602
  [\href{http://arXiv.org/abs/1310.3832}{{\tt 1310.3832}}].

\bibitem{Ammon:2016szz}
M.~Ammon, J.~Leiber and R.~P. Macedo, {\it {Phase diagram of 4D field theories
  with chiral anomaly from holography}},  {\em JHEP} {\bf 03} (2016) 164
  [\href{http://arXiv.org/abs/1601.02125}{{\tt 1601.02125}}].

\bibitem{Nicolis:2013lma}
A.~Nicolis, R.~Penco and R.~A. Rosen, {\it {Relativistic Fluids, Superfluids,
  Solids and Supersolids from a Coset Construction}},  {\em Phys. Rev.} {\bf
  D89} (2014), no.~4 045002 [\href{http://arXiv.org/abs/1307.0517}{{\tt
  1307.0517}}].

\bibitem{Son:2002sd}
D.~T. Son and A.~O. Starinets, {\it {Minkowski space correlators in AdS / CFT
  correspondence: Recipe and applications}},  {\em JHEP} {\bf 09} (2002) 042
  [\href{http://arXiv.org/abs/hep-th/0205051}{{\tt hep-th/0205051}}].

\bibitem{Baggioli:2014roa}
M.~Baggioli and O.~Pujolas, {\it {Electron-Phonon Interactions, Metal-Insulator
  Transitions, and Holographic Massive Gravity}},  {\em Phys. Rev. Lett.} {\bf
  114} (2015), no.~25 251602 [\href{http://arXiv.org/abs/1411.1003}{{\tt
  1411.1003}}].

\bibitem{Baggioli:2015gsa}
M.~Baggioli and D.~K. Brattan, {\it {Drag phenomena from holographic massive
  gravity}},  {\em Class. Quant. Grav.} {\bf 34} (2017), no.~1 015008
  [\href{http://arXiv.org/abs/1504.07635}{{\tt 1504.07635}}].

\bibitem{Alberte:2015isw}
L.~Alberte, M.~Baggioli, A.~Khmelnitsky and O.~Pujolas, {\it {Solid Holography
  and Massive Gravity}},  {\em JHEP} {\bf 02} (2016) 114
  [\href{http://arXiv.org/abs/1510.09089}{{\tt 1510.09089}}].

\bibitem{Argurio:2016xih}
R.~Argurio, G.~Giribet, A.~Marzolla, D.~Naegels and J.~A. Sierra-Garcia, {\it
  {Holographic Ward identities for symmetry breaking in two dimensions}},  {\em
  JHEP} {\bf 04} (2017) 007 [\href{http://arXiv.org/abs/1612.00771}{{\tt
  1612.00771}}].

\bibitem{Amoretti:2016bxs}
A.~Amoretti, D.~Areán, R.~Argurio, D.~Musso and L.~A. Pando~Zayas, {\it {A
  holographic perspective on phonons and pseudo-phonons}},  {\em JHEP} {\bf 05}
  (2017) 051 [\href{http://arXiv.org/abs/1611.09344}{{\tt 1611.09344}}].

\bibitem{Grozdanov:2015qia}
S.~Grozdanov, A.~Lucas, S.~Sachdev and K.~Schalm, {\it {Absence of
  disorder-driven metal-insulator transitions in simple holographic models}},
  {\em Phys. Rev. Lett.} {\bf 115} (2015), no.~22 221601
  [\href{http://arXiv.org/abs/1507.00003}{{\tt 1507.00003}}].

\bibitem{Alberte:2016xja}
L.~Alberte, M.~Baggioli and O.~Pujolas, {\it {Viscosity bound violation in
  holographic solids and the viscoelastic response}},  {\em JHEP} {\bf 07}
  (2016) 074 [\href{http://arXiv.org/abs/1601.03384}{{\tt 1601.03384}}].

\bibitem{Jokela:2017ltu}
N.~Jokela, M.~Jarvinen and M.~Lippert, {\it {Holographic pinning}},
  \href{http://arXiv.org/abs/1708.07837}{{\tt 1708.07837}}.

\bibitem{Alberte:2017cch}
L.~Alberte, M.~Ammon, M.~Baggioli, A.~Jiménez and O.~Pujolàs, {\it {Black
  hole elasticity and gapped transverse phonons in holography}},
  \href{http://arXiv.org/abs/1708.08477}{{\tt 1708.08477}}.

\bibitem{Headrick:2009pv}
M.~Headrick, S.~Kitchen and T.~Wiseman, {\it {A New approach to static
  numerical relativity, and its application to Kaluza-Klein black holes}},
  {\em Class. Quant. Grav.} {\bf 27} (2010) 035002
  [\href{http://arXiv.org/abs/0905.1822}{{\tt 0905.1822}}].

\bibitem{Adam:2011dn}
A.~Adam, S.~Kitchen and T.~Wiseman, {\it {A numerical approach to finding
  general stationary vacuum black holes}},  {\em Class. Quant. Grav.} {\bf 29}
  (2012) 165002 [\href{http://arXiv.org/abs/1105.6347}{{\tt 1105.6347}}].

\bibitem{Wiseman:2011by}
T.~Wiseman, {\em {Numerical construction of static and stationary black
  holes}}.
\newblock 2011.

\bibitem{Hartnoll:2009sz}
S.~A. Hartnoll, {\it {Lectures on holographic methods for condensed matter
  physics}},  {\em Class. Quant. Grav.} {\bf 26} (2009) 224002
  [\href{http://arXiv.org/abs/0903.3246}{{\tt 0903.3246}}].

\bibitem{Bagrov:2016cnr}
A.~Bagrov, N.~Kaplis, A.~Krikun, K.~Schalm and J.~Zaanen, {\it {Holographic
  fermions at strong translational symmetry breaking: a Bianchi-VII case
  study}},  {\em JHEP} {\bf 11} (2016) 057
  [\href{http://arXiv.org/abs/1608.03738}{{\tt 1608.03738}}].

\bibitem{Henningson:1998gx}
M.~Henningson and K.~Skenderis, {\it {The Holographic Weyl anomaly}},  {\em
  JHEP} {\bf 07} (1998) 023 [\href{http://arXiv.org/abs/hep-th/9806087}{{\tt
  hep-th/9806087}}].

\bibitem{deHaro:2000vlm}
S.~de~Haro, S.~N. Solodukhin and K.~Skenderis, {\it {Holographic reconstruction
  of space-time and renormalization in the AdS / CFT correspondence}},  {\em
  Commun. Math. Phys.} {\bf 217} (2001) 595--622
  [\href{http://arXiv.org/abs/hep-th/0002230}{{\tt hep-th/0002230}}].

\bibitem{Lucas:2015lna}
A.~Lucas, {\it {Hydrodynamic transport in strongly coupled disordered quantum
  field theories}},  {\em New J. Phys.} {\bf 17} (2015), no.~11 113007
  [\href{http://arXiv.org/abs/1506.02662}{{\tt 1506.02662}}].

\bibitem{Donos:2012gg}
A.~Donos and J.~P. Gauntlett, {\it {Helical superconducting black holes}},
  {\em Phys. Rev. Lett.} {\bf 108} (2012) 211601
  [\href{http://arXiv.org/abs/1203.0533}{{\tt 1203.0533}}].

\bibitem{Donos:2017mhp}
A.~Donos, J.~P. Gauntlett, T.~Griffin, N.~Lohitsiri and L.~Melgar, {\it
  {Holographic DC conductivity and Onsager relations}},  {\em JHEP} {\bf 07}
  (2017) 006 [\href{http://arXiv.org/abs/1704.05141}{{\tt 1704.05141}}].

\bibitem{Davison:2015taa}
R.~A. Davison, B.~Goutéraux and S.~A. Hartnoll, {\it {Incoherent transport in
  clean quantum critical metals}},  {\em JHEP} {\bf 10} (2015) 112
  [\href{http://arXiv.org/abs/1507.07137}{{\tt 1507.07137}}].

\bibitem{Martin:1972}
P.~C. Martin, O.~Parodi and P.~S. Pershan, {\it Unified hydrodynamic theory for
  crystals, liquid crystals, and normal fluids},  {\em Phys. Rev. A} {\bf 6}
  (Dec, 1972) 2401--2420.

\bibitem{Zippelius:1980}
A.~Zippelius, B.~I. Halperin and D.~R. Nelson, {\it Dynamics of two-dimensional
  melting},  {\em Phys. Rev. B} {\bf 22} (Sep, 1980) 2514--2541.

\bibitem{Delacretaz:2014jka}
L.~V. Delacr\'etaz, A.~Nicolis, R.~Penco and R.~A. Rosen, {\it {Wess-Zumino
  Terms for Relativistic Fluids, Superfluids, Solids, and Supersolids}},  {\em
  Phys. Rev. Lett.} {\bf 114} (2015), no.~9 091601
  [\href{http://arXiv.org/abs/1403.6509}{{\tt 1403.6509}}].

\bibitem{Nicolis:2015sra}
A.~Nicolis, R.~Penco, F.~Piazza and R.~Rattazzi, {\it {Zoology of condensed
  matter: Framids, ordinary stuff, extra-ordinary stuff}},  {\em JHEP} {\bf 06}
  (2015) 155 [\href{http://arXiv.org/abs/1501.03845}{{\tt 1501.03845}}].

\bibitem{Kadanoff}
L.~P. {Kadanoff} and P.~C. {Martin}, {\it {Hydrodynamic equations and
  correlation functions}},  {\em Annals of Physics} {\bf 24} (Oct., 1963)
  419--469.

\bibitem{Hartnoll:2007ih}
S.~A. Hartnoll, P.~K. Kovtun, M.~Muller and S.~Sachdev, {\it {Theory of the
  Nernst effect near quantum phase transitions in condensed matter, and in
  dyonic black holes}},  {\em Phys. Rev.} {\bf B76} (2007) 144502
  [\href{http://arXiv.org/abs/0706.3215}{{\tt 0706.3215}}].

\bibitem{Kovtun:2012rj}
P.~Kovtun, {\it {Lectures on hydrodynamic fluctuations in relativistic
  theories}},  {\em J. Phys.} {\bf A45} (2012) 473001
  [\href{http://arXiv.org/abs/1205.5040}{{\tt 1205.5040}}].

\bibitem{Hernandez:2017mch}
J.~Hernandez and P.~Kovtun, {\it {Relativistic magnetohydrodynamics}},  {\em
  JHEP} {\bf 05} (2017) 001 [\href{http://arXiv.org/abs/1703.08757}{{\tt
  1703.08757}}].

\bibitem{Grozdanov:2017kyl}
S.~Grozdanov and N.~Poovuttikul, {\it {Generalised global symmetries and
  magnetohydrodynamic waves in a strongly interacting holographic plasma}},
  \href{http://arXiv.org/abs/1707.04182}{{\tt 1707.04182}}.

\bibitem{Blake:2015epa}
M.~Blake, {\it {Momentum relaxation from the fluid/gravity correspondence}},
  {\em JHEP} {\bf 09} (2015) 010 [\href{http://arXiv.org/abs/1505.06992}{{\tt
  1505.06992}}].

\bibitem{Blake:2015hxa}
M.~Blake, {\it {Magnetotransport from the fluid/gravity correspondence}},  {\em
  JHEP} {\bf 10} (2015) 078 [\href{http://arXiv.org/abs/1507.04870}{{\tt
  1507.04870}}].

\bibitem{Burikham:2016roo}
P.~Burikham and N.~Poovuttikul, {\it {Shear viscosity in holography and
  effective theory of transport without translational symmetry}},  {\em Phys.
  Rev.} {\bf D94} (2016), no.~10 106001
  [\href{http://arXiv.org/abs/1601.04624}{{\tt 1601.04624}}].

\bibitem{Lucas:2015sya}
A.~Lucas, J.~Crossno, K.~C. Fong, P.~Kim and S.~Sachdev, {\it {Transport in
  inhomogeneous quantum critical fluids and in the Dirac fluid in graphene}},
  {\em Phys. Rev.} {\bf B93} (2016), no.~7 075426
  [\href{http://arXiv.org/abs/1510.01738}{{\tt 1510.01738}}].

\bibitem{Banks:2016krz}
E.~Banks, A.~Donos, J.~P. Gauntlett, T.~Griffin and L.~Melgar, {\it
  {Holographic thermal DC response in the hydrodynamic limit}},  {\em Class.
  Quant. Grav.} {\bf 34} (2017), no.~4 045001
  [\href{http://arXiv.org/abs/1609.08912}{{\tt 1609.08912}}].

\bibitem{Scopelliti:2017sga}
V.~Scopelliti, K.~Schalm and A.~Lucas, {\it {Hydrodynamic charge and heat
  transport on inhomogeneous curved spaces}},
  \href{http://arXiv.org/abs/1705.04325}{{\tt 1705.04325}}.

\bibitem{Hartnoll:2014lpa}
S.~A. Hartnoll, {\it {Theory of universal incoherent metallic transport}},
  {\em Nature Phys.} {\bf 11} (2015) 54
  [\href{http://arXiv.org/abs/1405.3651}{{\tt 1405.3651}}].

\bibitem{Davison:2014lua}
R.~A. Davison and B.~Goutéraux, {\it {Momentum dissipation and effective
  theories of coherent and incoherent transport}},  {\em JHEP} {\bf 01} (2015)
  039 [\href{http://arXiv.org/abs/1411.1062}{{\tt 1411.1062}}].

\bibitem{Mathematica10}
{Wolfram Research, Inc.}, {\em Mathematica, Version 10.2}.
\newblock Champaign, Illinois, 2015.

\end{thebibliography}\endgroup

\end{document}